\documentclass[journal]{IEEEtran}
\usepackage{graphicx}
\usepackage{amsmath}
\usepackage{multirow}
\usepackage{amssymb}
\usepackage{amsmath}
\usepackage{multirow}
\usepackage{colortbl}
\usepackage{tabularx}
\usepackage{tabulary}
\usepackage{tabularray}
\usepackage{hyperref}
\usepackage{algorithm}
\usepackage{algorithmic}
\usepackage{amsmath}
\usepackage{comment}
\usepackage{hyperref}       
\usepackage{url}            
\usepackage{booktabs}       
\usepackage{amsfonts}       
\usepackage{nicefrac}       
\usepackage{microtype}      
\usepackage[table,xcdraw]{xcolor}         
\usepackage{graphicx}
\usepackage{amsmath}
\usepackage{ulem}

\usepackage{dsfont}
\hypersetup{
    colorlinks=false,
    linkcolor=blue,
    filecolor=magenta,      
    urlcolor=cyan,
}

\ifCLASSINFOpdf
\else
\fi
\begin{document}
%
\title{Toward Foundational Model for Sleep Analysis  Using a Multimodal Hybrid Self-Supervised Learning Framework}
%
%
%

\author{Cheol-Hui Lee, 
        Hakseung Kim,
        Byung C. Yoon
        and Dong-Joo Kim

\IEEEcompsocitemizethanks{
    \IEEEcompsocthanksitem Cheol-Hui Lee is with the Department of Brain and Cognitive Engineering, Korea University, Seoul, South Korea; with Interdisciplinary Program in Precision Public Health, Korea University, Seoul, South Korea
    \IEEEcompsocthanksitem Hakseung Kim is with the Department of Brain and Cognitive Engineering, Korea University, Seoul, South Korea
    \IEEEcompsocthanksitem Byung C. Yoon is with the Department of Radiology, Stanford University School of Medicine, VA Palo Alto Health Care System, Palo Alto, CA, USA
    \IEEEcompsocthanksitem Dong-Joo Kim is with the Department of Brain and Cognitive Engineering, Korea University, Seoul, South Korea; with the Department of Neurology, Korea University College of Medicine, Seoul, South Korea; with Interdisciplinary Program in Precision Public Health, Korea University, Seoul, South Korea

    E-mail: dongjookim@korea.ac.kr}
}


%
%

\markboth{Journal of \LaTeX\ Class Files,~Vol.~14, No.~8, August~2015}%
{Shell \MakeLowercase{\textit{et al.}}: Bare Demo of IEEEtran.cls for IEEE Journals}
%



\maketitle

\begin{abstract}
Sleep is essential for maintaining human health and quality of life. Analyzing physiological signals during sleep is critical in assessing sleep quality and diagnosing sleep disorders. However, manual diagnoses by clinicians are time-intensive and subjective. Despite advances in deep learning that have enhanced automation, these approaches remain heavily dependent on large-scale labeled datasets. This study introduces SynthSleepNet, a multimodal hybrid self-supervised learning framework designed for analyzing polysomnography (PSG) data. SynthSleepNet effectively integrates masked prediction and contrastive learning to leverage complementary features across multiple modalities, including electroencephalogram (EEG), electrooculography (EOG), electromyography (EMG), and electrocardiogram (ECG). This approach enables the model to learn highly expressive representations of PSG data. Furthermore, a temporal context module based on Mamba was developed to efficiently capture contextual information across signals. \textbf{SynthSleepNet achieved superior performance compared to state-of-the-art methods across three downstream tasks}: sleep-stage classification, apnea detection, and hypopnea detection, with accuracies of 89.89\%, 99.75\%, and 89.60\%, respectively. The model demonstrated robust performance in a semi-supervised learning environment with limited labels, achieving accuracies of 87.98\%, 99.37\%, and 77.52\% in the same tasks. These results underscore the potential of the model as a foundational tool for the comprehensive analysis of PSG data. SynthSleepNet demonstrates comprehensively superior performance across multiple downstream tasks compared to other methodologies, making it expected to set a new standard for sleep disorder monitoring and diagnostic systems. The source code is available at https://github.com/dlcjfgmlnasa/SynthSleepNet.
\end{abstract}

\begin{IEEEkeywords}
foundation model, multimodal self-supervised learning, polysomnography, automatic sleep staging, AHI detection
\end{IEEEkeywords}

%
\IEEEpeerreviewmaketitle

\section{Introduction}

Sleep is critical in maintaining human health, alleviating mental and physical stress, and preserving physiological balance \cite{ref1}. Many individuals experience sleep disorders, prompting clinicians to use polysomnography (PSG) for diagnosing and monitoring physiological changes during sleep. PSG records multiple physiological signals, including electroencephalograms (EEG), electrooculograms (EOG), electromyograms (EMG), electrocardiograms (ECG), and airflow signals, providing comprehensive insights into sleep-related activities \cite{ref2, ref3}. Based on these recorded signals, clinicians perform various sleep-related diagnoses and assessments, which often require labor-intensive manual interpretation of complex physiological signals and a highly specialized expertise \cite{ref3}. For instance, accurate sleep staging involves identifying subtle EEG patterns such as K-complexes and sleep spindles \cite{camilleri2014automatic}, while detecting sleep apnea requires precise analysis of respiratory effort and airflow changes \cite{bahammam2011evaluation}.

Researchers have proposed various deep-learning-based algorithms for automated sleep assessment to address these challenges. However, most of these approaches rely on supervised learning, which demands large amounts of labeled data \cite{ref4, ref5, ref6, ref7, ref8, ref9, ref10, ref11, ref12, ref13, ref14, phyo2022transsleep}. Recently, self-supervised learning (SSL), a method for extracting meaningful representations from unlabeled data, has been applied to PSG data \cite{ref15, ref16, ref17, ref18, ref19, ref20, ref21, ref22, ref23, ko2024eeg, zhang2024brant}. SSL facilitates the discovery of high-level semantic patterns without labels by training on pseudo-labels generated through predefined tasks, creating a generalized network that can be fine-tuned for specific downstream applications. SSL methodologies are broadly classified as single-modality \cite{ref15, ref16, ref17, ref18, ref19, ko2024eeg} and multimodal approaches \cite{ref20, ref21, ref22, ref23, zhang2024brant}, with the latter integrating information from multiple modalities.

Existing methodologies have demonstrated significant advancements but retain notable limitations. Most studies are designed for single tasks, primarily focusing on sleep stage classification \cite{ref13, ref15, ref16, ref17, ref18, ref19, ref20, ref21, ref22}. However, clinicians evaluate sleep quality using multiple indicators, rendering these approaches overly restrictive \cite{ref2, ref3, ref24}. In particular, detecting apnea and hypopnea is crucial for assessing various health risks, including cardiovascular \cite{drager2017sleep} and neurological disorders \cite{ferini2017neurological}. However, the transient and variable nature of these events, combined with the scarcity of large, well-annotated datasets, presents substantial challenges to developing generalized models. Additionally, current methodologies predominantly employ contrastive learning within the SSL paradigm. While effective in optimizing inter-modality relationships and learning discriminative representations, this approach has drawbacks. However, this approach is highly sensitive to the quality of data augmentation, and poorly designed or inconsistent augmentation strategies can significantly degrade model performance. This issue is particularly pronounced in complex and highly variable data, such as physiological signals \cite{ref17, ref19}.

This study proposes SynthSleepNet, a multimodal hybrid SSL methodology for analyzing PSG data. Drawing inspiration from NeuroNet \cite{ref19} and MultiMAE \cite{ref27}, SynthSleepNet introduces a novel multimodal SSL framework that combines masked prediction and contrastive learning to jointly exploit generative and discriminative representation learning. This combination enables SynthSleepNet to overcome the individual limitations of each method—preserving semantic context without masking-induced instability, while simultaneously enforcing representational separability across physiological events. Moreover, this integration allows robust cross-modal fusion under low supervision by aligning modality-specific features using jointly optimized reconstruction and contrastive objectives. Experimental results demonstrate that this design improves generalization under limited labels and stabilizes pretraining dynamics. Unlike existing methodologies, SynthSleepNet evaluates sleep quality comprehensively by performing three downstream tasks: sleep stage classification, apnea detection, and hypopnea detection. Experimental results demonstrate that SynthSleepNet surpasses state-of-the-art methods across all three tasks and excels in semi-supervised learning environments. The proposed methodology is expected to establish a new foundation for sleep analysis.

\section{Related Work}
\subsection{Self-Supervised Learning Methodology for Sleep Assessment with Single-Modal Physiological Signals}

SSL methodologies designed for single-modal physiological signals primarily target sleep-stage classification, focusing predominantly on EEG. BENDR \cite{ref15} incorporates a convolution neural network (CNN)-based module to extract EEG features and a transformer to capture temporal contexts across signals. This model employs contrastive learning by designing output vectors from the CNN-based module and transformer as positive pairs if they correspond to the same time point while treating others as negative pairs. ContraWR \cite{ref16} replaces the standard InfoNCE loss used in contrastive learning with triplet loss, which minimizes and maximizes the distances between positive and negative pairs, respectively. In this framework, negative pairs are defined as the mean of each sample. TS-TCC \cite{ref17} applies two distinct augmentations to the same EEG data and utilizes a temporal contrasting module to enhance similarity between the contexts of identical samples while reducing similarity between different contexts of distinct samples. Similarly, mulEEG \cite{ref18} drops the augmentation methodology of TS-TCC \cite{ref17} but extends it using multiview SSL to improve learning. This approach incorporates EEG signals and spectrograms as input data, leveraging a diverse loss function to extract complementary information across multiple views. TRIPNet \cite{ko2024eeg} utilizes three pretext tasks—topped band prediction, spatial noise recognition, and temporal trend identification—to capture the spectral, spatial, and temporal characteristics of EEG signals. Additionally, it incorporates a statistician module to address inter-subject and inter-condition variability. Multi-view SSL for sleep EEG was proposed in \cite{yu2025multi}, which separately trains temporal and spectral views using dedicated encoders and combines intra-view contrastive loss with cross-view contrastive loss. Furthermore, it employs dynamic weighting between losses to achieve balanced learning across views. NeuroNet \cite{ref19} introduces an integrated approach combining masked-prediction-based SSL with contrastive-learning-based SSL to derive unique and discriminative representations. Employing a masked autoencoder structure, NeuroNet \cite{ref19} performs masked prediction while simultaneously processing two differently sampled vectors through an encoder. The network optimizes learning using the NT-Xent loss for contrastive learning, enhancing its ability to identify meaningful patterns and representations in EEG data.

\subsection{Multimodal Self-Supervised Learning Methodology for Sleep Assessment with Multimodal Physiological Signals}

Several multimodal SSL methodologies have been designed for sleep-stage classification, leveraging multiple physiological signal modalities. MVCC \cite{ref20} incorporates an intra-view temporal contrastive module to extract temporal features within individual modalities and an inter-view consistency contrastive module to ensure coherence across multiple signal modalities. COCOA \cite{ref21} introduces a cross-modality correlation loss to maximize the similarity between representations of different modalities for the same sample while minimizing the similarity between representations of different time intervals within the same modality. This is achieved using an intra-modality discriminator loss, which refines representation quality. CroSSL \cite{ref22} is distinguished by its robust flexibility, particularly in scenarios with missing data. The method employs the VICReg loss to minimize the dissimilarity between representations of different modalities. Brant-X \cite{zhang2024brant} employs patch-level and sequence-level alignment modules based on InfoNCE to enable consistent representation learning across multiple physiological signals. Patch-level alignment ensures consistency between signals at the same time point, while sequence-level alignment maintains semantic consistency across different time points. SleepFM \cite{ref23} adopts a leave-one-out contrastive learning strategy based on the InfoNCE loss and applies it to various sleep-related downstream tasks. MVCC \cite{ref20}, COCOA \cite{ref21}, CroSSL \cite{ref22}, Brant-X \cite{zhang2024brant} and SleepFM \cite{ref23} represent contrastive-learning-based multimodal SSL methodologies.

\begin{figure*}[!t]
\centering
\includegraphics[width=1\textwidth]{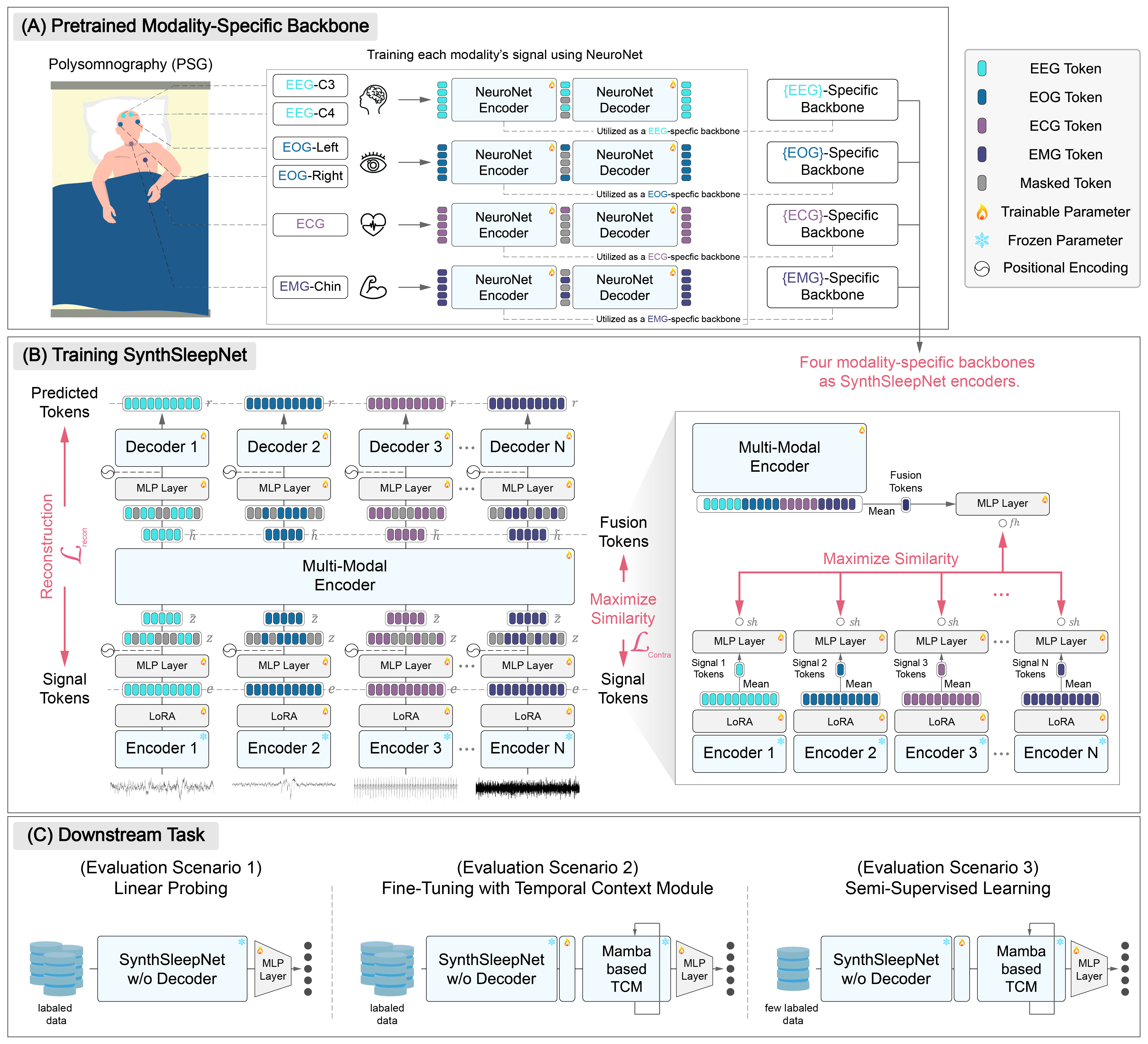}
\caption{Overall architecture. (A) Training process of the modality-specific backbone, which extracts features from physiological signals for each modality. The pretrained backbone serves as the encoder for SynthSleepNet. (B) Training workflow of SynthSleepNet—a multimodal hybrid self-supervised learning framework. (C) The pretrained SynthSleepNet (excluding the decoder) is applied to three evaluation scenarios: linear probing, fine-tuning with temporal context module, and semi-supervised learning.}
\vspace{-5.5mm}
\label{fig:figure1}
\end{figure*}

\section{Methodology}

SynthSleepNet introduces an advanced multimodal hybrid SSL framework to comprehensively integrate diverse physiological signaling modalities. Figure \ref{fig:figure1} illustrates the overall architecture of SynthSleepNet, with detailed explanations provided below.

\subsection{SynthSleepNet: Multimodal Hybrid Self-Supervised Learning Framework for PSG}

\subsubsection{Model Architecture} ~

\textit{(Modality-Specific Backbone)} The modality-specific backbone was tailored to extract features unique to each physiological signal modality. It is based on the NeuroNet \cite{ref19} architecture, which is an SSL methodology that combines the ability to identify unique information from data samples through masked prediction tasks with the discriminative representation capabilities provided by contrastive learning tasks. In this study, physiological signals with similar characteristics were grouped, learned using NeuroNet, and then utilized as modality-specific backbones. To elaborate, after grouping physiological signals with similar characteristics, each group was learned using NeuroNet, with the encoder of NeuroNet serving as the modality-specific backbone. Ultimately, the four modality-specific backbones learned from EEG, EOG, EMG, and ECG were integrated as key components within the SynthSleepNet architecture. This process is illustrated in Figure \ref{fig:figure1} (A).

\textit{(Encoder)} The encoder serves as a foundational component of SynthSleepNet, mirroring the structure of the modality-specific backbone. SynthSleepNet integrates pretrained modality-specific backbones into encoders that are optimized for the unique attributes of the input physiological signals. When these modality-specific backbones are integrated, the model parameters are frozen to preserve the learned representations. For instance, when processing EEG C4 \& C3 channels and EOG Left \& Right channels, SynthSleepNet incorporates ``pretrained EEG-specific backbones'' and ``pretrained EOG-specific backbones,'' assigning two encoders for each signal type. LoRA \cite{ref28} was applied to each encoder to enhance the precision of information extraction. LoRA \cite{ref28} facilitates efficient fine-tuning by employing rank-decomposition weight matrices, thereby minimizing the need to alter the entire weight set. This process facilitates each encoder to produce output vectors $\{e_i^m\}_{i=1}^N$, representing the ``signal tokens'' illustrated in Figure 1 (B), where $N$ indicates the number of tokens, $m \in [1,2,3,\dots,M]$ identifies the index of the input physiological signal, and $M$ denotes the total number of input physiological signals, leading to $M$ encoders.

\textit{(Multimodal Encoder)} The multimodal encoder integrates modality-specific features extracted by individual encoders using a standard Vision Transformer (ViT) \cite{ref29}. The input to the encoder consisted of tokens generated through a three-step process. First, the output vectors $\{e_i^m\}_{i=1}^N$ from each encoder were passed through separate projection layers, and positional encoding was added. This resulted in a new set of vectors, $\{z_i^m\}_{i=1}^N$. Next, a subset of these vectors, $\{z_i^m\}_{i=1}^N$, was randomly sampled, while the remaining tokens were masked. This sampled subset is denoted by $\{\tilde{z}_i^m\}_{i=1}^{\tilde{N}}$, where $\tilde{N}$ represents the number of sampled tokens. Finally, the sampled output vectors, $\{\tilde{z}_i^m\}_{i=1}^{\tilde{N}}$, from all encoders were concatenated and fed into the multimodal encoder. This produced the output vectors $\{\{\tilde{h}_i^m\}_{i=1}^{\tilde{N}}\}_{m=1}^M$, which represent the ``fusion tokens'' (see Figure \ref{fig:figure1} (B)).

\textit{(Decoder)} Separate ViT decoders were used for each representation vector of physiological signal to reconstruct the masked tokens, resulting in a total of $M$ decoders corresponding to the number of encoders. Decoders also used ViT \cite{ref29}. However, the decoders were removed after the SSL phase. The decoder considered the output vectors $\{\{\tilde{h}_i^m\}_{i=1}^{\tilde{N}}\}_{m=1}^M$ as input from the multimodal encoder corresponding to each physiological signal combined with the masked vectors. After that, the input vectors were processed through a projection layer and positional encoding. The masked vector represented the vectors excluded during random sampling and contained information omitted from the input data. Each decoder generated $\{d_i^m\}_{i=1}^N$ through this process, corresponding to the ``predicted tokens'' (see Figure \ref{fig:figure1} (B)).

\subsubsection{Training Objectives} ~

\textit{(Masked Prediction)} Masked prediction involved concealing specific portions of the input data and training the model to predict the hidden parts. This approach enabled the model to infer missing information and learn intrinsic patterns and relationships within the data. The mean square error (MSE) loss was applied to the masked prediction. Specifically, the output vectors $\{d_i^m\}_{i=1}^N$ produced by each decoder were passed through a projection layer to obtain transformed vectors $\{r_i^m\}_{i=1}^N$, which matched the size of the output vectors $\{e_i^m\}_{i=1}^N$ of the encoder (the output vectors of the encoder represent the ``signal tokens'' in Figure \ref{fig:figure1} (B)). After that, the MSE loss was computed between $\{r_i^m\}_{i=1}^N$ and $\{e_i^m\}_{i=1}^N$, focusing solely on the masked vectors. The loss function is expressed as:

\begin{equation}
L_{\text{recon}} = \frac{1}{M(N-\tilde{N})} \sum_{m=1}^M \sum_{i=1}^{N-\tilde{N}} (e_i^m - r_i^m)^2
\end{equation} where $M$ represents the number of physiological signals, $N$ denotes the total number of tokens, and $\tilde{N}$ corresponds to the number of sampled tokens. This approach ensured that the model focused on accurately reconstructing the masked portions of the input data.

\textit{(Contrastive Learning)} SynthSleepNet incorporated the NT-Xent loss \cite{ref30} to optimize the relationships between the output vectors of the encoder and the multimodal encoder. Specifically, the method reduces the distance between output vectors for identical inputs to both the encoder and the multimodal encoder. Conversely, it increases the distance for different input instances. This alignment ensures that the semantic information extracted by the encoder for individual signals is consistent with the semantic information extracted by the multimodal encoder when processing all signals together, enabling effective integration of information from multiple physiological signals.

Encoder output vectors $\{e_i^m\}_{i=1}^N$ were averaged elementwise to derive $\mathit{signal}^m$. Similarly, multimodal encoder output vectors $\{\{\tilde{h}_i^m\}_{i=1}^{\tilde{N}}\}_{m=1}^M$ were averaged to produce the $\mathit{fusion}$ representation. After that, these representations, $\{\mathit{signal}^m\}_{m=1}^M$ and $\mathit{fusion}$, were mapped to a latent space and normalized, resulting in $\{\mathit{sh}^m\}_{m=1}^M$ and $\mathit{fh}$, respectively. The NT-Xent loss \cite{ref30} was applied as follows:

\begin{equation}
L_{\text{contra}} = \frac{1}{2NM} \sum_{m=1}^M \sum_{k=1}^N \left[ l(2k-1, 2k, m) + l(2k, 2k-1, m) \right]
\end{equation}

\begin{equation}
l(i, j, m) = -\log \left( 
\frac{\exp\left(\mathit{sim}(\mathit{fh}_i, \mathit{sh}_j^m)/\tau\right)}
{\sum_{k=1}^{2N} 1_{[k \neq i]} \exp\left(\mathit{sim}(\mathit{fh}_i, \mathit{sh}_k^m)/\tau\right)}
\right)
\end{equation} here, $N$ represents the batch size, $\mathit{sim}$ refers to cosine similarity, and $\tau > 0$ corresponds to the temperature scaling factor.

\textit{(Joint Loss)} 
SynthSleepNet combined masked prediction and contrastive learning to generate robust representations of physiological signals. Masked prediction captured semantic features through reconstruction, while contrastive learning aligned relationships across individual and multimodal signal representations. The combined loss function is expressed as:
\begin{equation}
L_{\text{total}} = L_{\text{recon}} + \alpha L_{\text{contra}}
\end{equation}
where $\alpha$ denotes a balancing hyperparameter that controls the relative contribution of the two loss components ($L_{\text{recon}}$ and $L_{\text{contra}}$).

\begin{figure}[!b]
    \centerline{\includegraphics[width=\columnwidth]{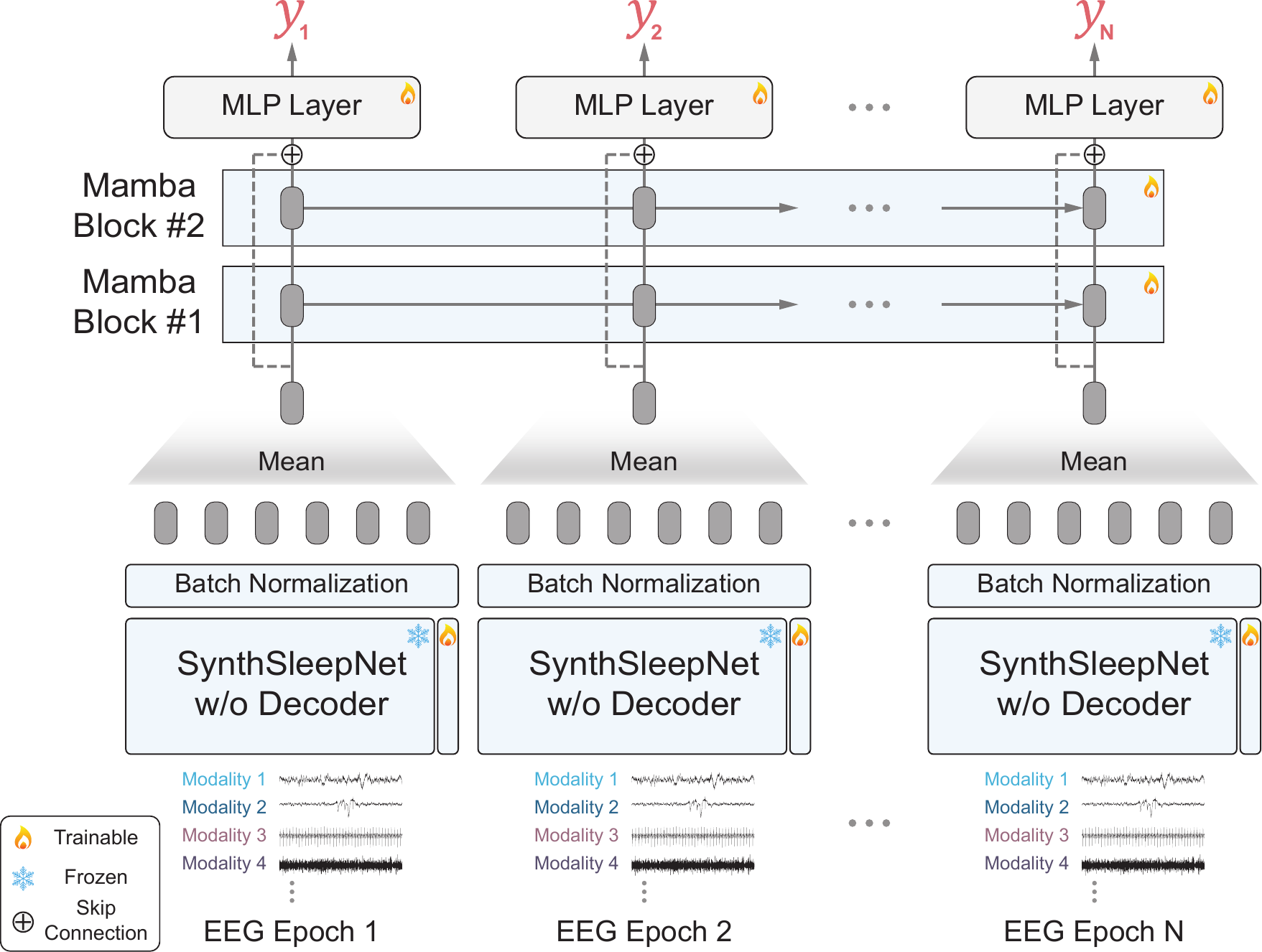}}
    \caption{Structure of the Mamba-based temporal context module.}
    \label{fig:figure2}
    \vspace{-5mm}
\end{figure}

\subsection{Mamba based on Temporal Context Module}

The American Academy of Sleep Medicine (AASM) \cite{ref24} guidelines emphasize that sleep stage classification relies on local features within individual PSG epochs and relationships between adjacent epochs. For example, the AASM smoothing rule \cite{ref24} identifies sleep stages—such as Wake, REM, or Non-REM—that persist for 3–5 minutes or longer as new cycles. Shorter stages are often treated as transient and disregarded. Consequently, an effective sleep-stage classification model requires a module capable of capturing inter-epoch features across multiple PSG epochs. This module is referred to as the temporal context module (TCM). Most existing studies implement TCMs using recurrent neural networks (RNNs) or multihead attention mechanisms. However, this study introduces a novel TCM model based on the Mamba \cite{ref31}, a linear-time sequence modeling approach.

Mamba \cite{ref31} was designed to address the efficiency challenges associated with conventional sequence models like LSTMs and Transformers. While LSTMs can capture both short- and long-term dependencies, their reliance on gating mechanisms can result in issues such as vanishing gradients when handling long sequences \cite{al2023lstm}. Meanwhile, Transformers excel at modeling long-range dependencies through self-attention, but their quadratic computational complexity limits their efficiency for longer sequences \cite{keles2023computational}. To overcome these challenges, Mamba \cite{ref31} introduces a State-Space Model (SSM)-based framework that selectively emphasizes or suppresses different parts of the input sequence. This selective mechanism enables Mamba \cite{ref31} to efficiently capture complex temporal patterns in physiological signals, without the computational overhead of self-attention or the stability issues associated with recurrent architectures.

As a structured SSM, Mamba \cite{ref31} operates as a continuous system, mapping an input function \( x(t) \) to a high-dimensional latent state \( h(t) \), which is then used to generate the output \( y(t) \). This latent state contains \( F \) feature channels and evolves continuously over time according to the following equations:
\begin{equation}
\dot{h}(t) = A h(t) + B x(t), \quad y(t) = C^T h(t)
\end{equation} where \( A \in \mathbb{R}^{F \times F} \) and \( B \in \mathbb{R}^{F \times 1} \) represent the evolution parameter and the input transformation parameter, respectively, while \( C \in \mathbb{R}^{F \times 1} \) is the projection parameter.

However, as structured SSMs rely on continuous-time mapping, they cannot be directly applied to discrete sequences. To overcome this, zero-order holding is used for discretization, allowing the model to sample data at discrete time steps \( \Delta \). The discretized state-space representation is then given by:
\begin{equation}
\bar{A} = \exp(\Delta A), \quad \bar{B} = (\Delta A)^{-1} (\exp(\Delta A) - I) \cdot \Delta B
\end{equation}
\begin{equation}
h(t) = \bar{A} h(t-1) + \bar{B} x(t), \quad y(t) = C^T h(t)
\end{equation} where \( (\bar{A}, \bar{B}, C) \) are dynamically adjusted within the selective state-space modeling framework, similar to the gating mechanism in RNNs. This mechanism enables Mamba \cite{ref31} to dynamically emphasize or suppress specific parts of the input sequence, resulting in more efficient and robust sequence modeling.

The Mamba-based TCM was inspired by the IITNet-style TCM architecture \cite{ref13}. The proposed Mamba-based TCM involved the following six steps: (1) The pretrained SynthSleepNet was modified by removing its decoder, resulting in the SynthSleepNet w/o decoder, which served as the backbone network. This network extracted a vector sequence $F = \{f_i\}_{i=1}^N$, corresponding to a single PSG epoch. (2) The vector sequence $F$ was averaged element-wise to produce token $K$. (3) The backbone network for $T$ PSG epochs processed each epoch individually to extract vector sequences. These sequences were averaged element-wise to produce the vectors $\{K_i\}_{i=1}^T$, which were batch-normalized to stabilize learning. (4) The vectors $\{K_i\}_{i=1}^T$ were input into the Mamba model to temporal dependencies across PSG epochs, resulting in output vectors. (5) The input vectors $\{K_i\}_{i=1}^T$ and $\{M_i\}_{i=1}^T$ were summed element-wise (i.e., skip connections). (6) The combined vectors were passed through an projection layer to produce the final output vector. Figure \ref{fig:figure2} is an illustration of the Mamba-based TCM.

\begin{table*}[t]
\renewcommand{\arraystretch}{1.2}
\centering
\caption{Modalities and downstream tasks used in previous approaches.}
\label{table:table1}

\scalebox{0.95}{
{\scriptsize
\begin{tabular}{c|c|c|c} 
\hline
\textit{Model Name}  & \textit{Year} & \textit{Modality}   & \textit{Downstream Task}                                \\ 
\hline
\multicolumn{4}{c}{supervised learning + single-modality}                                                            \\ 
\hline
IITNet \cite{ref13}      & 2020          & EEG                 & sleep stage classification                              \\
AttnSleep \cite{ref4}    & 2021          & EEG                 & sleep stage classification                              \\
SleepExpertNet \cite{ref5}  & 2023          & EEG                 & sleep stage classification                              \\
TransSleep \cite{phyo2022transsleep}       & 2023          & EEG                 & sleep stage classification                                         \\ 
RAFNet \cite{ref7}       & 2023          & ECG                 & apnea detection                                         \\ 
\hline
\multicolumn{4}{c}{supervised learning + multi-modality}                                                             \\ 
\hline
U-Sleep \cite{ref8}      & 2021          & EEG,
  EOG          & sleep stage classification                              \\
SalientSleepNet \cite{ref9} & 2021          & EEG, EOG            & sleep stage classification                              \\
XSleepNet \cite{ref10}   & 2022          & EEG,
  EOG, EMG     & sleep stage classification                              \\
MMASleepNet \cite{ref11} & 2022          & EEG, EOG, EMG       & sleep stage classification                              \\
DynamicSleepNet \cite{ref12} & 2023          & EEG, EOG, EMG       & sleep stage classification                              \\
BioSig-UNet \cite{ref14}     & 2024          & EEG, EMG, ECG, Resp & arousal detection                                       \\ 
\hline
\multicolumn{4}{c}{SSL + single-modality}                                                                            \\ 
\hline
BENDR \cite{ref15}           & 2021          & EEG                 & sleep stage classification                              \\
ContraWR \cite{ref16}        & 2023          & EEG                 & sleep stage classification                              \\
TS-TCC \cite{ref17}          & 2021          & EEG                 & sleep stage classification                              \\
mulEEG \cite{ref18}          & 2022          & EEG                 & sleep stage classification                              \\
NeuroNet \cite{ref19}        & 2024          & EEG                 & sleep stage classification                              \\ 
\hline
\multicolumn{4}{c}{SSL + multi-modality}                                                                             \\ 
\hline
MVCC \cite{ref20}            & 2023          & EEG, EOG            & sleep stage classification                              \\
COCOA \cite{ref21}           & 2022          & EEG, EOG, EMG       & sleep stage classification                              \\
CroSSL \cite{ref22}          & 2024          & EEG, EOG, EMG       & sleep stage classification                              \\
SleepFM \cite{ref23}         & 2024          & EEG, EOG, ECG, Resp & sleep stage classification, sleep-disordered detection  \\ 
\hline
\multicolumn{4}{r}{* Resp = Respiration sign.}  \\     
\multicolumn{4}{r}{}

\end{tabular}}
}
\vspace{-10mm}
\end{table*}
\section{Experiments}
\subsection{Dataset Description and Data Preprocessing}
The Sleep Heart Health Study (SHHS) \cite{ref33} is a multicenter cohort study aimed at examining the cardiovascular and other health outcomes associated with sleep-disordered breathing. The dataset consisted of two subsets: SHHS1 and SHHS2. Each subset included PSG recordings of multiple physiological signals, specifically two bipolar EEG channels (C4-A1 and C3-A2), one ECG channel, two EOG channels (Left, Right), two leg EMG channels, a snore sensor, pulse oximeters, and a body position sensor. Two EEG channels, two EOG channels, one ECG channel, and one leg EMG channel were selected for analysis. These signals underwent the following preprocessing steps: (1) All physiological signals were resampled to 100 Hz. (2) A robust scaler optimized for physiological data was applied to reduce the influence of outliers while preserving the relative scale of the features. (3) Different bandpass filters were applied according to the signal type: 0.5–40 Hz, 3–30 Hz, and 25–50 Hz for the EEG and EOG channels, ECG channel, and EMG channel, respectively. Only data from SHHS1 cells were used for this study.

\textit{(Sleep Stage Classification)} Each 30-s segment in the dataset was annotated by sleep experts into one of eight categories: Wake, Non-REM1 (N1), Non-REM2 (N2), Non-REM3 (N3), Non-REM4 (N4), REM, Movement, and Unknown. N3 and N4 were merged into a single class (N3), while the "Movement" and "Unknown" categories were excluded to adhere to the AASM standard.

\textit{(Apnea Detection)} The SHHS dataset included annotations for three types of apnea events: obstructive, central, and mixed apnea. These categories were consolidated into a single class. Each 30-s segment was labeled as 1 if an apnea event was detected and 0 otherwise.

\textit{(Hypopnea Detection)} Hypopnea-related events in the SHHS dataset were recorded as a single category. Similarly, each 30-s segment was labeled as 1 if a hypopnea event was present or 0 otherwise.

\subsection{Evaluation Schema}
Five-fold subject-group cross-validation was conducted to assess the performance of the methodologies. The evaluation framework was tailored for both SSL-based methodologies and supervised methodologies. The dataset was divided into three subsets: pre-training, training, and testing, with a distribution ratio of 7:2:1. The pre-trained group was used for unsupervised training of the SSL model without labels. The training group, consisting of a small subset of labeled data, was used for linear evaluation and fine-tuning. It involved attaching a downstream classifier to the SSL-trained network and completing downstream tasks. The test group was used for the final performance evaluation. The dataset was split into training and test subsets for supervised methodologies, with the test subset remaining consistent across both SSL-based and supervised methodologies. Three evaluation scenarios were implemented (see Figure \ref{fig:figure1} (c)). Detailed descriptions of these scenarios are provided below.

\textit{(Evaluation Scenario 1: Linear Probing)} The backbone network (i.e., SynthSleepNet w/o decoder) remained fixed in this scenario, while only the downstream classifier was trained. The evaluation aimed to determine the effectiveness of each SSL methodology in capturing representations of PSG data.

\textit{(Evaluation Scenario 2: Fine-Tuning with Temporal Context Module)} This scenario evaluated the combined model, SynthSleepNet+TCM, which integrated the backbone network with the TCM. Most parameters of the backbone network were frozen except for a specific segment of the final layer in the multimodal encoder (i.e., the attention projection layer). This approach facilitated additional learning of nonlinear features in the data and leveraged multi-epoch information, which was expected to outperform Scenario 1. Additionally, this evaluation method compared SSL-based methodologies and supervised learning approaches.

\textit{(Evaluation Scenario 3: Semi-Supervised Learning)} This scenario examined the performance of semi-supervised learning methods. The proposed methodologies, SynthSleepNet and SynthSleepNet+TCM, were compared with SalientSleepNet (a supervised learning-based methodology) and SleepFM (an SSL-based methodology). Adjustments were made to ensure the number of modalities was consistent across all methods. Only approximately 1\% and 5\% of the labeled data from the entire dataset were used for training for the semi-supervised learning experiments. 

Appendix Tables 1, 2, and 3 list the hyperparameter values used in each evaluation scenario.

\subsection{Performance Metrics}
Three metrics were employed to evaluate the performance of the proposed model: overall accuracy (\textit{ACC}), macro F1 score (\textit{MF1}), and Cohen’s kappa coefficient (\textit{Kappa}). These metrics are widely recognized, with \textit{MF1} and \textit{Kappa} being particularly suitable for datasets with class imbalances. The formulae for \textit{ACC} and \textit{MF1} are as follows:
\begin{equation}
\textit{ACC} = \frac{\sum_{i=1}^K TP_i}{N}
\end{equation}
\begin{equation}
\textit{MF1} = \frac{\sum_{i=1}^K F1_i}{K}
\end{equation}
where $TP_i$ and $N$ represent the true positives for the $i$-th class and the total number of samples, respectively. Similarly, $F1_i$ and $K$ denote the F1 score for the $i$-th class and the total number of classes, respectively. \textit{Kappa} measures the level of agreement between two observers on categorical values and is expressed as:
\begin{equation}
\textit{Kappa} \equiv \frac{p_o - p_e}{1 - p_e} = 1 - \frac{1 - p_o}{1 - p_e}
\end{equation}
where $p_o$ denotes the observed accuracy and $p_e$ represents the expected chance agreement.

\subsection{Other Approaches}

Representative methodologies were selected and implemented across four paradigms to evaluate the performances of SynthSleepNet. Five methodologies were chosen and implemented under \textit{“SSL + single-modality”}: BENDR \cite{ref15}, ContraWR \cite{ref16}, TS-TCC \cite{ref17}, mulEEG \cite{ref18}, and NeuroNet \cite{ref19}. Four methodologies were selected under \textit{“SSL + multi-modality”}: MVCC \cite{ref20}, COCOA \cite{ref21}, CroSSL \cite{ref22}, and SleepFM \cite{ref23}. Four methodologies were implemented in \textit{“supervised learning + single-modality”}: IITNet \cite{ref13}, AttnSleep \cite{ref4}, SleepExpertNet \cite{ref5}, TransSleep \cite{phyo2022transsleep} and RAFNet \cite{ref7}. Finally, five methodologies were selected and implemented under the paradigm of \textit{“supervised learning + multi-modality”}: U-Sleep \cite{ref8}, SalientSleepNet \cite{ref9}, XSleepNet \cite{ref10}, MMASleepNet \cite{ref11}, and DynamicSleepNet \cite{ref12}. Brief descriptions of these methodologies are provided below.

\begin{table*}[!htbp]
\caption{Performance comparison of linear probing with existing methodologies.}
\label{table:table2}
\centering
\resizebox{\textwidth}{!}{
\renewcommand{\arraystretch}{1.10}
{\scriptsize
\begin{tabular}{c|c|c|c|ccc|ccc|ccc} 
\hline
\multirow{4}{*}{\begin{tabular}[c]{@{}c@{}}\textit{Model}\\\textit{Name}\end{tabular}} & \multirow{4}{*}{\begin{tabular}[c]{@{}c@{}}\textit{Backbone}\\\textit{Architecture}\end{tabular}} & \multirow{4}{*}{\begin{tabular}[c]{@{}c@{}}\textit{SSL}\\\textit{Training}\\\textit{Type}\end{tabular}} & \multirow{4}{*}{\textit{Modality
  (Count)}} & \multicolumn{9}{c}{\textit{Performance}}                                                                                                                                                                                                                                                                           \\ 
\cline{5-13}
                                                                                       &                                                                                                   &                                                                                                         &                                     & \multicolumn{3}{c|}{\begin{tabular}[c]{@{}c@{}}\textit{Sleep Stage}\\\textit{Classification}\end{tabular}} & \multicolumn{3}{c|}{\begin{tabular}[c]{@{}c@{}}\textit{Apnea}\\\textit{Detection}\end{tabular}} & \multicolumn{3}{c}{\begin{tabular}[c]{@{}c@{}}\textit{Hypopnea}\\\textit{Detection}\end{tabular}}  \\ 
\cline{5-13}
                                                                                       &                                                                                                   &                                                                                                         &                                     & \textit{ACC}   & \textit{MF1}   & \textit{$K$}                                                               & \textit{ACC}   & \textit{MF1}   & \textit{$K$}                                                    & \textit{ACC}   & \textit{MF1}   & \textit{$K$}                                                        \\ 
\hline
BENDR \cite{ref15}                                                                                  & CNN + VIT                                                                                         & CL                                                                                                      & EEG1                                & 65.50          & 57.17          & 0.54                                                                     & 80.16          & 46.15          & 0.15                                                          & 77.10          & 59.98          & 0.25                                                              \\
ContraWR \cite{ref16}                                                                              & CNN                                                                                               & CL                                                                                                      & EEG1                                & 76.31          & 67.28          & 0.68                                                                     & 93.27          & 52.42          & 0.07                                                          & 72.48          & 55.98          & 0.19                                                              \\
TS-TCC \cite{ref17}                                                                                & CNN                                                                                               & CL                                                                                                      & EEG1                                & 73.48          & 64.39          & 0.64                                                                     & 73.57          & 43.46          & 0.03                                                          & 72.60          & 55.96          & 0.19                                                              \\
mulEEG \cite{ref18}                                                                                & CNN                                                                                               & CL                                                                                                      & EEG1                                & 75.35          & 66.36          & 0.66                                                                     & 78.79          & 45.49          & 0.10                                                          & 74.61          & 58.26          & 0.23                                                              \\
NeuroNet \cite{ref19}                                                                              & CNN
  + VIT                                                                                       & MP
  + CL                                                                                               & EEG1                                & 77.29          & 68.25          & 0.69                                                                     & 79.59          & 44.99          & 0.10                                                          & 73.04          & 56.49          & 0.19                                                              \\ 
\hline
MVCC \cite{ref20}                                                                                  & CNN                                                                                               & CL                                                                                                      & EEG1 + EOG1                         & 76.24          & 67.19          & 0.68                                                                     & 80.14          & 44.99          & 0.08                                                          & 59.46          & 45.78          & 0.06                                                              \\
COCOA \cite{ref21}                                                                                 & CNN                                                                                               & CL                                                                                                      & EEG1 + EOG1 + EMG1                  & 73.12          & 64.28          & 0.64                                                                     & 96.81          & 56.36          & 0.14                                                          & 78.76          & 62.24          & 0.29                                                              \\
CroSSL \cite{ref22}                                                                                & CNN                                                                                               & CL                                                                                                      & EEG2 + EOG1 + EMG1                  & 78.93          & 69.96          & 0.71                                                                     & 98.52          & 63.06          & 0.27                                                          & 78.61          & 61.99          & 0.28                                                              \\
SleepFM \cite{ref23}                                                                               & CNN                                                                                               & CL                                                                                                      & EEG2 + EOG2 + ECG1                  & 80.17          & 71.27          & 0.73                                                                     & 99.29          & 71.89          & 0.44                                                          & 79.14          & 62.50          & 0.29                                                              \\ 
\hline
\multirow{9}{*}{SynthSleepNet}                                                         & NeuroNet                                                                                          & MP
  + CL                                                                                               & EEG2                                & 80.03          & 70.00          & 0.73                                                                     & 97.82          & 49.88          & -0.02                                                         & 76.25          & 59.31          & 0.24                                                              \\
                                                                                       & NeuroNet                                                                                          & MP
  + CL                                                                                               & EOG2                                & 77.34          & 66.49          & 0.69                                                                     & 94.17          & 53.24          & 0.08                                                          & 76.04          & 59.34          & 0.24                                                              \\
                                                                                       & NeuroNet                                                                                          & MP
  + CL                                                                                               & EEG1 + EOG1                         & 81.22          & 71.33          & 0.74                                                                     & 96.61          & 57.57          & 0.16                                                          & 73.94          & 57.33          & 0.21                                                              \\
                                                                                       & NeuroNet                                                                                          & MP
  + CL                                                                                               & EEG2 + EOG2                         & 83.28          & 73.98          & 0.77                                                                     & 98.39          & 61.41          & 0.23                                                          & 74.34          & 57.65          & 0.21                                                              \\
                                                                                       & NeuroNet                                                                                          & MP
  + CL                                                                                               & EEG1 + EOG1 + EMG1                  & 82.62          & 74.09          & 0.64                                                                     & 99.31          & 72.56          & 0.45                                                          & 80.21          & 63.64          & 0.31                                                              \\
                                                                                       & NeuroNet                                                                                          & MP
  + CL                                                                                               & EEG1 + EOG1 + ECG1                  & 79.96          & 70.47          & 0.73                                                                     & \textbf{99.54} & 77.35          & 0.55                                                          & 80.70          & 64.65          & 0.33                                                              \\
                                                                                       & NeuroNet                                                                                          & MP
  + CL                                                                                               & EEG2 + EOG2 + EMG1                  & \textbf{84.36} & \uline{75.29}  & \textbf{0.78}                                                            & 99.40          & 74.50          & 0.49                                                          & 80.86          & 64.62          & 0.33                                                              \\
                                                                                       & NeuroNet                                                                                          & MP
  + CL                                                                                               & EEG2 + EOG2 + ECG1                  & 81.77          & 72.66          & 0.75                                                                     & 99.43          & \uline{78.22}  & \uline{0.56}                                                  & \uline{83.77}  & \uline{67.95}  & \uline{0.38}                                                      \\
                                                                                       & NeuroNet                                                                                          & MP
  + CL                                                                                               & EEG2 + EOG2 + EMG1 +ECG1            & \uline{83.23}  & \textbf{75.36} & \uline{0.77}                                                             & \uline{99.46}  & \textbf{78.96} & \textbf{0.58}                                                 & \textbf{87.97} & \textbf{73.24} & \textbf{0.48}                                                     \\
\hline
\multicolumn{13}{r}{* CL = contrastive learning, MP = masked prediction, EEG1 = C4-A1 channel, EOG1 = EOG-Left channel} \\
\multicolumn{13}{r}{* The \textbf{best results} in each row are shown in bold, while the \uline{second-best} results are underlined // $K$ = \textit{Kappa}} \\
\end{tabular}}
}
\vspace{-4mm}
\end{table*}
\begin{table*}[!htbp]
\caption{Performance comparison of fine-tuning with temporal context module across methodologies.}
\label{table:table3}
\centering
\resizebox{\textwidth}{!}{
\renewcommand{\arraystretch}{1.10}
{\scriptsize
\begin{tabular}{c|c|c|ccc|ccc|ccc} 
\hline
\multirow{4}{*}{\begin{tabular}[c]{@{}c@{}}\textit{Model}\\\textit{Name}\end{tabular}}                                            & \multirow{4}{*}{\begin{tabular}[c]{@{}c@{}}\textit{Training}\\\textit{Type}\end{tabular}} & \multirow{4}{*}{\textit{Modality (Count)}} & \multicolumn{9}{c}{\textit{Performance}}                                                                                                                                                                                                                              \\ 
\cline{4-12}
                                                                               &                                        &                                            & \multicolumn{3}{c|}{\begin{tabular}[c]{@{}c@{}} \textit{Sleep Stage} \\\textit{Classification}\end{tabular}} & \multicolumn{3}{c|}{\begin{tabular}[c]{@{}c@{}}\textit{Apnea}\\\textit{Detection}\end{tabular}} & \multicolumn{3}{c}{\begin{tabular}[c]{@{}c@{}}\textit{Hypopnea}\\\textit{Detection}\end{tabular}}  \\ 
\cline{4-12}
                                                                               &                                        &                                            & \textit{ACC}            & \textit{MF1}            & \textit{K}                                                       & \textit{ACC}   & \textit{MF1}   & \textit{K}                                           & \textit{ACC}   & \textit{MF1}   & \textit{K}                                              \\ 
\hline
IITNet \cite{ref13}                                                                        & Supervised                             & EEG1                                       & 83.55          & 76.00          & 0.77                                                    & 57.12          & 44.00          & 0.04                                        & 71.15          & 55.11          & 0.18                                           \\
AttnSleep \cite{ref4}                                                                     & Supervised                             & EEG1                                       & 81.38          & 72.49          & 0.74                                                    & 62.94          & 48.59          & 0.09                                        & 71.17          & 54.96          & 0.17                                           \\
SleepExpertNet \cite{ref5}                                                                & Supervised                             & EEG1                                       & 84.33          & 76.93          & 0.79                                                    & 62.76          & 48.59          & 0.10                                        & 71.00          & 54.96          & 0.17                                           \\
TransSleep \cite{phyo2022transsleep}                                                                        & Supervised                             & EEG1                                       & 81.31          & 75.44          & 0.74                                                    & 58.33          & 42.81          & 0.05                                        & 70.96          & 52.45          & 0.17                                           \\ 

RAFNet \cite{ref7}                                                                        & Supervised                             & ECG1                                       & 48.50          & 42.49          & 0.33                                                    & 99.56          & 77.84          & 0.56                                        & 88.95          & 74.96          & 0.51                                           \\ 
\hline
U-Sleep \cite{ref8}                                                                       & Supervised                             & EEG1
  + EOG1                              & 86.56          & 79.98          & 0.82                                                    & 66.76          & 51.59          & 0.13                                        & 75.35          & 59.10          & 0.24                                           \\
SalientSleepNet \cite{ref9}                                                               & Supervised                             & EEG1 + EOG1                                & 88.62          & 82.47          & 0.84                                                    & 65.93          & 51.18          & 0.13                                        & 75.19          & 58.47          & 0.22                                           \\
XSleepNet \cite{ref10}                                                                     & Supervised                             & EEG1 + EOG1 + EMG1                         & 87.60          & 80.99          & 0.83                                                    & 98.69          & 64.20          & 0.29                                        & 77.68          & 61.36          & 0.27                                           \\
MMASleepNet \cite{ref11}                                                                   & Supervised                             & EEG1 + EOG1 + EMG1                         & 86.65          & 79.99          & 0.82                                                    & 98.82          & 65.60          & 0.31                                        & 75.66          & 59.34          & 0.24                                           \\
DynamicSleepNet \cite{ref12}                                                               & Supervised                             & EEG1 + EOG1 + EMG1                         & 87.06          & 80.50          & 0.82                                                    & 98.75          & 65.09          & 0.30                                        & 78.76          & 62.16          & 0.28                                           \\
BioSig-UNet \cite{ref14}                                                                   & Supervised                             & EEG2 + EMG1 + ECG1                         & 76.11              & 64.14              & 0.53                                                       & 82.33              & 53.52              & 0.18                                           & 76.95              & 49.34              & 0.20                                              \\ 
\hline
\multirow{9}{*}{\begin{tabular}[c]{@{}c@{}}SynthSleepNet\\+\\TCM\end{tabular}} & SSL                                    & EEG2                                       & 87.34          & 81.58          & 0.82                                                    & 83.95          & 47.00          & 0.02                                        & 76.45          & 60.21          & 0.25                                           \\
                                                                               & SSL                                    & EOG2                                       & 85.15          & 78.41          & 0.80                                                    & 96.96          & 56.75          & 0.15                                        & 77.17          & 60.53          & 0.26                                           \\
                                                                               & SSL                                    & EEG1
  + EOG1                              & 88.31          & 82.01          & 0.84                                                    & 96.95          & 56.78          & 0.15                                        & 75.75          & 59.23          & 0.24                                           \\
                                                                               & SSL                                    & EEG2
  + EOG2                              & 89.21          & 83.75          & 0.85                                                    & 98.38          & 62.43          & 0.25                                        & 75.82          & 59.21          & 0.24                                           \\
                                                                               & SSL                                    & EEG1 + EOG1 + EMG1                         & 89.28          & 83.47          & 0.85                                                    & 99.44          & 75.18          & 0.50                                        & 81.68          & 65.77          & 0.35                                           \\
                                                                               & SSL                                    & EEG1 + EOG1 + ECG1                         & 88.17          & 82.65          & 0.84                                                    & 99.64          & 80.43          & 0.61                                        & 88.53          & 74.07          & 0.49                                           \\
                                                                               & SSL                                    & EEG2 + EOG2 + EMG1                         & \textbf{89.89} & \textbf{84.52} & \textbf{0.86}                                           & 99.37          & 73.46          & 0.47                                        & 83.08          & 67.20          & 0.37                                           \\
                                                                               & SSL                                    & EEG2 + EOG2 + ECG1                         & 88.76          & 83.03          & 0.84                                                    & \uline{99.69}  & \uline{82.36}  & \uline{0.65}                                & \textbf{89.60} & \textbf{75.96} & \textbf{0.53}                                  \\
                                                                               & SSL                                    & EEG2 + EOG2 + EMG1 +ECG1                   & \uline{89.61}  & \uline{83.98}  & \uline{0.86}                                            & \textbf{99.75} & \textbf{84.53} & \textbf{0.69}                               & \uline{89.32}  & \uline{75.47}  & \uline{0.52}                                   \\
\hline
\multicolumn{12}{r}{* EEG1 = C4-A1 channel, EOG1 = EOG-Left channel} \\
\multicolumn{12}{r}{* The \textbf{best results} in each row are shown in bold, while the \uline{second-best} results are underlined // $K$ = \textit{Kappa}} \\
\end{tabular}}
}
\vspace{-5mm}
\end{table*}
\begin{table*}[!htbp]
\caption{Performance comparison of semi-supervised learning across methodologies.}
\label{table:table4}
\centering
\resizebox{\textwidth}{!}{
\renewcommand{\arraystretch}{1.10}
{\tiny 
\begin{tabular}{c|c|c|ccc|ccc|ccc} 
\hline
\multicolumn{12}{c}{\textit{1\% of labeled data}}                                                                                                                                                                                                                                                                                                                                                                                                                                                                                                     \\ 
\hline
\multirow{2}{*}{\begin{tabular}[c]{@{}c@{}}\textit{Model}\\\textit{Name}\end{tabular}}                                                   & \multirow{2}{*}{\begin{tabular}[c]{@{}c@{}}\textit{Training}\\\textit{Type}\end{tabular}} & \multirow{2}{*}{\textit{Modality (Count)}} & \multicolumn{3}{c|}{\begin{tabular}[c]{@{}c@{}}\textit{Sleep Stage}\\\textit{Classification}\end{tabular}} & \multicolumn{3}{c|}{\begin{tabular}[c]{@{}c@{}}\textit{Apnea}\\\textit{Detection}\end{tabular}} & \multicolumn{3}{c}{\begin{tabular}[c]{@{}c@{}}\textit{Hypopnea}\\\textit{Detection}\end{tabular}}  \\ 
\cline{4-12}
                                                                                       &                                                                                            &                                            & \textit{ACC}   & \textit{MF1}   & \textit{K}                                                                & \textit{ACC}   & \textit{MF1}   & \textit{K}                                                    & \textit{ACC}   & \textit{MF1}   & \textit{K}                                                       \\ 
\hline
SalientSleepNet \cite{ref9}                                                                       & Supervised                                                                                 & EEG1
  + EOG1                              & 61.44          & 53.49          & 0.49                                                                      & 79.34          & 45.32          & 0.02                                                          & 50.22          & 39.65          & 0.01                                                             \\
SleepFM \cite{ref23}                                                                                & SSL                                                                                        & EEG1
  + EOG1                              & 70.14          & 61.49          & 0.60                                                                      & 93.48          & 52.02          & 0.07                                                          & 50.95          & 39.77          & 0.00                                                             \\
SynthSleepNet                                                                          & SSL                                                                                        & EEG1
  + EOG1                              & 78.04          & 65.61          & 0.70                                                                      & 95.39          & 54.08          & 0.10                                                          & 63.87          & 49.33          & 0.10                                                             \\
SynthSleepNet+TCM                                                                      & SSL                                                                                        & EEG1
  + EOG1                              & \uline{84.71}  & \uline{76.48}  & \uline{0.79}                                                              & 95.78          & 54.64          & 0.11                                                          & 65.93          & 50.82          & 0.12                                                             \\ 
\hline
SalientSleepNet \cite{ref9}                                                                       & Supervised                                                                                 & EEG1
  + EOG1 + EMG1                       & 60.75          & 52.90          & 0.48                                                                      & 83.64          & 46.93          & 0.02                                                          & 50.34          & 39.50          & 0.00                                                             \\
SleepFM \cite{ref23}                                                                               & SSL                                                                                        & EEG1
  + EOG1 + EMG1                       & 71.73          & 62.89          & 0.62                                                                      & 96.51          & 55.98          & 0.13                                                          & 55.07          & 42.49          & 0.02                                                             \\
SynthSleepNet                                                                          & SSL                                                                                        & EEG1
  + EOG1 + EMG1                       & 76.80          & 65.48          & 0.68                                                                      & 98.88          & 66.60          & 0.33                                                          & 69.98          & 54.32          & 0.17                                                             \\
SynthSleepNet+TCM                                                                      & SSL                                                                                        & EEG1
  + EOG1 + EMG1                       & \textbf{85.01} & \textbf{78.34} & \textbf{0.79}                                                             & 99.09          & 68.96          & 0.38                                                          & 74.05          & 57.44          & 0.21                                                             \\ 
\hline
SalientSleepNet \cite{ref9}                                                                       & Supervised                                                                                 & EEG1
  + EOG1 + ECG1                       & 48.68          & 42.60          & 0.33                                                                      & 78.10          & 44.80          & 0.01                                                          & 55.23          & 43.16          & 0.04                                                             \\
SleepFM  \cite{ref23}                                                                              & SSL                                                                                        & EEG1
  + EOG1 + ECG1                       & 66.16          & 60.20          & 0.56                                                                      & 94.68          & 53.21          & 0.09                                                          & 60.08          & 46.46          & 0.07                                                             \\
SynthSleepNet                                                                          & SSL                                                                                        & EEG1
  + EOG1 + ECG1                       & 77.29          & 65.13          & 0.69                                                                      & \uline{99.23}  & \uline{71.00}  & \uline{0.42}                                                  & \uline{75.40}  & \uline{59.03}  & \uline{0.24}                                                     \\
SynthSleepNet+TCM                                                                      & SSL                                                                                        & EEG1
  + EOG1 + ECG1                       & 83.77          & 75.74          & 0.78                                                                      & \textbf{99.35} & \textbf{72.94} & \textbf{0.46}                                                 & \textbf{76.93} & \textbf{60.34} & \textbf{0.25}                                                    \\ 
\hline
\multicolumn{12}{c}{\textit{5\% of labeled data}}                                                                                                                                                                                                                                                                                                                                                                                                                                                                                                     \\ 
\hline
\multirow{2}{*}{\begin{tabular}[c]{@{}c@{}}\textit{Model}\\\textit{Name}\end{tabular}} & \multirow{2}{*}{\begin{tabular}[c]{@{}c@{}}\textit{Training}\\\textit{Type}\end{tabular}}  & \multirow{2}{*}{\textit{Modality (Count)}} & \multicolumn{3}{c|}{\begin{tabular}[c]{@{}c@{}}\textit{Sleep Stage}\\\textit{Classification}\end{tabular}} & \multicolumn{3}{c|}{\begin{tabular}[c]{@{}c@{}}\textit{Apnea}\\\textit{Detection}\end{tabular}} & \multicolumn{3}{c}{\begin{tabular}[c]{@{}c@{}}\textit{Hypopnea}\\\textit{Detection}\end{tabular}}  \\ 
\cline{4-12}
                                                                                       &                                                                                            &                                            & \textit{ACC}   & \textit{MF1}   & \textit{K}                                                                & \textit{ACC}   & \textit{MF1}   & \textit{K}                                                    & \textit{ACC}   & \textit{MF1}   & \textit{K}                                                       \\ 
\hline
SalientSleepNet \cite{ref9}                                                                       & Supervised                                                                                 & EEG1
  + EOG1                              & 67.30          & 58.88          & 0.56                                                                      & 88.89          & 49.20          & 0.04                                                          & 50.81          & 40.00          & 0.01                                                             \\
SleepFM  \cite{ref23}                                                                              & SSL                                                                                        & EEG1
  + EOG1                              & 71.87          & 62.90          & 0.62                                                                      & 95.00          & 53.53          & 0.09                                                          & 51.13          & 40.14          & 0.01                                                             \\
SynthSleepNet                                                                          & SSL                                                                                        & EEG1
  + EOG1                              & 80.50          & 69.29          & 0.73                                                                      & 95.80          & 54.64          & 0.01                                                          & 62.78          & 48.55          & 0.09                                                             \\
SynthSleepNet+TCM                                                                      & SSL                                                                                        & EEG1
  + EOG1                              & \textbf{87.98} & \textbf{82.12} & \textbf{0.83}                                                             & 96.89          & 56.80          & 0.15                                                          & 67.73          & 52.33          & 0.14                                                             \\ 
\hline
SalientSleepNet \cite{ref9}                                                                       & Supervised                                                                                 & EEG1
  + EOG1 + EMG1                       & 66.87          & 58.40          & 0.55                                                                      & 90.31          & 50.00          & 0.05                                                          & 51.41          & 40.32          & 0.01                                                             \\
SleepFM  \cite{ref23}                                                                              & SSL                                                                                        & EEG1
  + EOG1 + EMG1                       & 73.27          & 64.31          & 0.64                                                                      & 97.98          & 60.42          & 0.21                                                          & 59.11          & 45.82          & 0.06                                                             \\
SynthSleepNet                                                                          & SSL                                                                                        & EEG1
  + EOG1 + EMG1                       & 80.92          & 71.13          & 0.74                                                                      & 99.07          & 68.72          & 0.38                                                          & 71.02          & 55.02          & 0.18                                                             \\
SynthSleepNet+TCM                                                                      & SSL                                                                                        & EEG1
  + EOG1 + EMG1                       & \uline{87.41}  & \uline{81.92}  & \uline{0.83}                                                              & \uline{99.27}  & \uline{71.70}  & \uline{0.44}                                                  & 74.54          & 58.27          & 0.22                                                             \\ 
\hline
SalientSleepNet \cite{ref9}                                                                        & Supervised                                                                                 & EEG1
  + EOG1 + ECG1                       & 51.67          & 45.53          & 0.37                                                                      & 85.95          & 47.92          & 0.03                                                          & 72.97          & 56.37          & 0.19                                                             \\
SleepFM \cite{ref23}                                                                               & SSL                                                                                        & EEG1
  + EOG1 + ECG1                       & 68.93          & 62.50          & 0.60                                                                      & 96.12          & 55.08          & 0.12                                                          & 61.41          & 47.49          & 0.08                                                             \\
SynthSleepNet                                                                          & SSL                                                                                        & EEG1
  + EOG1 + ECG1                       & 79.79          & 69.08          & 0.72                                                                      & 99.27          & 71.66          & 0.43                                                          & \uline{76.80}  & \uline{60.02}  & \uline{0.25}                                                     \\
SynthSleepNet+TCM                                                                      & SSL                                                                                        & EEG1
  + EOG1 + ECG1                       & 83.60          & 75.73          & 0.77                                                                      & \textbf{99.37} & \textbf{73.47} & \textbf{0.47}                                                 & \textbf{77.52} & \textbf{61.32} & \textbf{0.27}                                                    \\
\hline
\multicolumn{12}{r}{* EEG1 = C4-A1 channel, EOG1 = EOG-Left channel} \\
\multicolumn{12}{r}{* The \textbf{best results} in each row are shown in bold, while the \uline{second-best} results are underlined // $K$ = \textit{Kappa}} \\
\end{tabular}}
}
\vspace{-5mm}
\end{table*}

Table \ref{table:table1} summarizes all methodologies considered in this study. Most selected methods are designed for sleep-stage classification, which has been more extensively studied than other sleep assessment methodologies. Moreover, additional sleep assessment approaches include RAFNet \cite{ref7}, MultiUNet, and SleepFM \cite{ref23}, designed to detect apnea, arousal, and sleep disorders, respectively. Notably, SleepFM \cite{ref23} differs from other methodologies by performing two downstream tasks. Both BioSig-UNet \cite{ref14} and SleepFM \cite{ref23} use respiratory signals. However, these signals were excluded from the present study due to their limited quantity and poor quality in the SHHS dataset. Consequently, BioSig-UNet and SleepFM \cite{ref23} were appropriately reimplemented to align with the requirements of the study.

\section{Results}
\subsection{Evaluation Scenario 1: Linear Probing}

Single-modality SSL methodologies, including BENDR \cite{ref15}, ContraWR \cite{ref16}, TS-TCC \cite{ref17}, mulEEG \cite{ref18}, and NeuroNet \cite{ref19}, achieved effective results for sleep-stage classification but performed poorly in detecting apnea and hypopnea (see Table \ref{table:table2}). Specifically, \textit{ACC} exceeded 70, yet the \textit{Kappa} coefficient remained at or below 0.25 in apnea and hypopnea detection, signifying suboptimal performance. Conversely, multimodal SSL methodologies, such as COCOA \cite{ref21}, CroSSL \cite{ref22}, and SleepFM \cite{ref23} performed consistently well across three downstream tasks. Among these, SleepFM \cite{ref23} achieved notably high performance. However, it underperformed relative to SynthSleepNet when identical modalities (EEG2, EOG2, and ECG1) were used. The respective differences in \textit{MF1} scores across the three downstream tasks were approximately 1.39, 6.33, and 5.45.

SynthSleepNet demonstrated the highest performance across all three downstream tasks, underscoring its superior capacity for data representation compared to existing SSL approaches. Notably, SynthSleepNet exhibited optimal performance with the modality combination of “EEG2+EOG2+EMG1” for sleep stage classification and “EEG2+EOG2+EMG1+ECG1” for apnea and hypopnea detection. An analysis of the impact of different modality combinations revealed that EEG, EOG, and EMG contributed positively to sleep-stage classification, while ECG and EMG significantly enhanced apnea and hypopnea detection, with ECG playing a particularly critical role. For instance, the inclusion of ECG in the combination “EEG1+EOG1+ECG1” significantly improved performance in apnea detection, evidenced by a sharp increase in the \textit{Kappa} by approximately 0.39 compared to “EEG1+EOG1.”

\subsection{Evaluation Scenario 2: Fine-tuning with Temporal Context Module}

The results from evaluation scenario 2, incorporating TCM, indicated that SynthSleepNet+TCM achieved superior performance across all metrics compared with state-of-the-art supervised learning methodologies \cite{ref4, ref5, ref7, ref8, ref9, ref10, ref11, ref12, ref13} (Table \ref{table:table3}). Notably, SynthSleepNet+TCM demonstrated exceptional performance with only limited labeled data, while supervised learning approaches required relatively large volumes of labeled data.

SynthSleepNet exhibited optimal performance with specific modality combinations for different tasks: “EEG2+EOG2+EMG1” for sleep-stage classification, “EEG2+EOG2+EMG1+ECG1” for apnea detection, and “EEG2+EOG2+ECG1” for hypopnea detection. Furthermore, SynthSleepNet+TCM significantly outperformed SynthSleepNet. For instance, SynthSleepNet+TCM demonstrated improvements in \textit{MF1} scores of approximately 8.62, 5.57, and 2.23 in sleep-stage classification, apnea detection, and hypopnea detection using the “EEG2+EOG2+EMG1+ECG1” modality combination, respectively, compared to SynthSleepNet. These findings underscore the efficacy of the Mamba-based TCM and the fine-tuning approach in enhancing the performance of the model.

\subsection{Evaluation Scenario 3: Semi-Supervised Learning}

Table \ref{table:table4} comprehensively analyzes methodologies applied in a semi-supervised learning context. Scenario 3 evaluated SalientSleepNet \cite{ref9} and SleepFM \cite{ref23} as benchmarks against SynthSleepNet. SalientSleepNet \cite{ref9} and SleepFM \cite{ref23} are supervised learning methods and SSL-based approaches, respectively. Additionally, the evaluation used only approximately 1\% and 5\% of the labeled data subsets, contrary to the full labeled dataset.

The performance comparison between models trained on fully and minimally labeled data revealed a significant drop for SalientSleepNet \cite{ref9}. For instance, SalientSleepNet exhibited reductions of approximately 27.18, 28.98, and 0.35 in \textit{ACC}, \textit{MF1}, and \textit{Kappa}, respectively, during sleep-stage classification with the modality combination “EEG1 + EOG1” when using 1\% of labeled data compared to the fully labeled dataset (Table \ref{table:table2} and \ref{table:table3}). SleepFM \cite{ref23} displayed comparatively better performance across both 1\% and 5\% labeled data subsets, achieving higher \textit{ACC}, \textit{MF1}, and \textit{Kappa} values than SalientSleepNet \cite{ref9}. However, its performance also diminished substantially compared to results derived from fully labeled data. SynthSleepNet and SynthSleepNet+TCM demonstrated exceptional robustness under limited labeled data conditions, maintaining consistently high performance across all metrics. For example, SynthSleepNet+TCM exhibited minimal reductions in \textit{MF1} in apnea detection with the modality combination “EEG1 + EOG1 + ECG1,” with decreases of only 7.49 and 6.96 for 1\% and 5\% of labeled data, respectively, compared to results obtained with fully labeled datasets. Under the same conditions, SynthSleepNet exhibited that \textit{MF1} decreased by 9.43 and 8.77 for 1\% and 5\% of labeled data, respectively. This indicates that SynthSleepNet demonstrates consistent performance relative to other methodologies, albeit not at the level achieved by SynthSleepNet+TCM.

\subsection{Ablation Experiments}

Ablation experiments were conducted to evaluate the performance of the proposed model. All experiments used the modality combination “EEG2 + EOG2,” with SSL training limited to 20 epochs to enhance experimental efficiency. The evaluation criterion was the performance in sleep stage classification. The optimal hyperparameters were determined based on the outcomes of these ablation experiments.

\subsubsection{Evaluation Scenario 1: Linear Probing} ~

\textit{(Masking Ratios)} SynthSleepNet achieved the highest \textit{MF1} of 73.91 at a masking ratio of 40\%. Figure \ref{fig:figure3} illustrates that SynthSleepNet without contrastive learning exhibited optimal performance at masking ratios between 50\% and 70\%, while SynthSleepNet without masked prediction performed optimally at masking ratios between 30\% and 50\%. When both tasks were applied concurrently, performance exceeded that observed with either task in isolation, demonstrating that tasks complemented and reinforced each other when combined.

\textit{(Decoder Depth and Width)} Table \ref{table:table5} shows that SynthSleepNet achieved the highest \textit{ACC} and \textit{Kappa} values when the decoder dimension and depth were 256 and 2, respectively. The highest \textit{MF1} was observed when the decoder dimension and depth were 256 and 3, respectively. However, variations based on decoder size were minimal. This study incorporated the decoder configuration yielding the highest \textit{MF1} (decoder dimension: 256; decoder depth: 3) for training SynthSleepNet.

\textit{(Loss Balance Scale)}
An ablation study was conducted to optimize the balance between NT-Xent and MSE losses. Values of $\alpha < 1.0$ indicate a greater emphasis on the masked prediction task, while values of $\alpha > 1.0$ signify a stronger focus on contrastive learning. Table \ref{table:table6} shows that the model achieves optimal performance across all metrics when $\alpha = 1$, indicating that equal weighting of the two losses yields the best results.

\subsubsection{Evaluation Scenario 2: Fine-tuning with Temporal Context Module} ~

\textit{(Temporal Context Module)} This study investigated the most effective TCM structure for integrating and analyzing information across multiple PSG epochs. In this context, the employed Mamba-based structure outperformed the LSTM \cite{ref13} and Multi-Head Attention \cite{ref4, ref5} structure used in previous research in terms of both performance and efficiency (Table \ref{table:table7} and \ref{table:table8}). Furthermore, an analysis of the Mamba’s performance across different context lengths revealed that a context length of 20 achieved the highest \textit{ACC} and \textit{Kappa} values (Table \ref{table:table7}).

\subsection{Rebound Point} 
\begin{table}[!htbp]
\caption{Performance of linear probing with different decoder dimensions and depths.}
\label{table:table5}
\centering
\resizebox{0.3\textwidth}{!}{
\renewcommand{\arraystretch}{1.2}
{\scriptsize 
\begin{tabular}{cc|c|cccc} 
\cline{2-6}
 & \multicolumn{2}{c|}{\textit{Decoder}} & \multicolumn{3}{c}{\textit{Performance}}        &   \\ 
\cline{2-6}
 & \textit{Dim}         & \textit{Depth} & \textit{ACC}   & \textit{MF1}   & \textit{K}    &   \\ 
\cline{2-6}
 & \multirow{4}{*}{192} & 1              & 82.51          & 73.14          & 0.76          &   \\
 &                      & 2              & 82.30          & 72.92          & 0.76          &   \\
 &                      & 3              & 82.57          & 72.99          & 0.76          &   \\
 &                      & 4              & 81.37          & 73.13          & 0.75          &   \\ 
\cline{2-6}
 & \multirow{4}{*}{256} & 1              & 82.36          & 72.76          & 0.76          &   \\
 &                      & 2              & \textbf{82.92} & 72.96          & \textbf{0.77} &   \\
 &                      & \textbf{3}     & 82.71          & \textbf{73.91} & 0.76          &   \\
 &                      & 4              & 82.13          & 72.94          & 0.76          &   \\ 
\cline{2-6}
 & \multirow{4}{*}{512} & 1              & 82.46          & 72.86          & 0.76          &   \\
 &                      & 2              & 82.07          & 72.93          & 0.76          &   \\
 &                      & 3              & 82.87          & 73.46          & 0.76          &   \\
 &                      & 4              & 82.90          & 73.51          & 0.76          &   \\
\cline{2-6}
\multicolumn{7}{r}{\footnotesize *\textbf{best results} in each row are shown in bold.} \\
\multicolumn{7}{r}{\footnotesize *$K$ = \textit{Kappa}} 
\end{tabular}}
}
\label{table5}
\vspace{-3.5mm}
\end{table}

\begin{figure}[!htbp]
\centering
\includegraphics[width=0.4\textwidth]{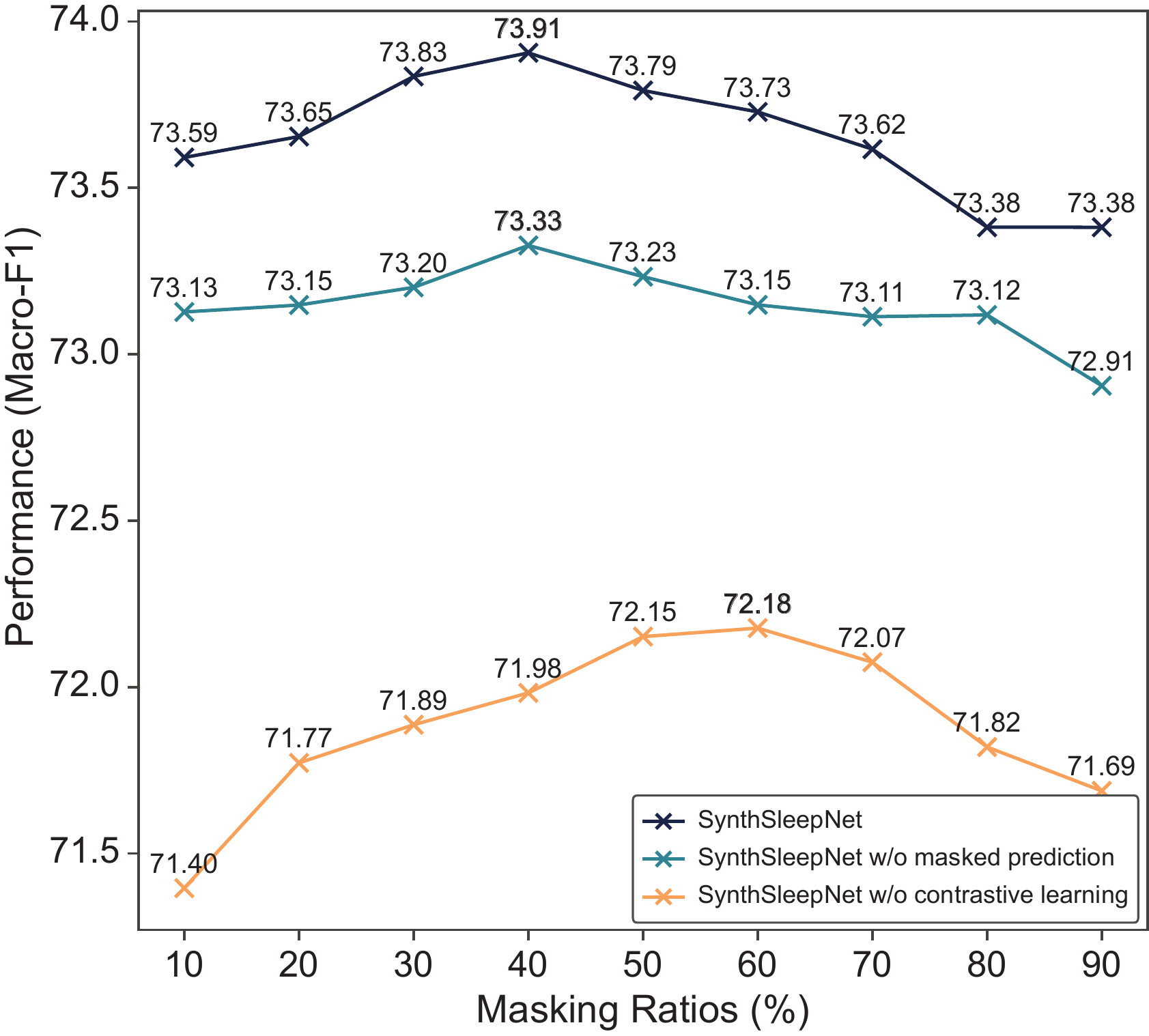}
\caption{Effect of different masking ratios on SynthSleepNet performance.}
\vspace{-4mm}
\label{fig:figure3}
\end{figure}

\begin{figure*}[!htbp]
\centering
\includegraphics[width=0.8\textwidth]{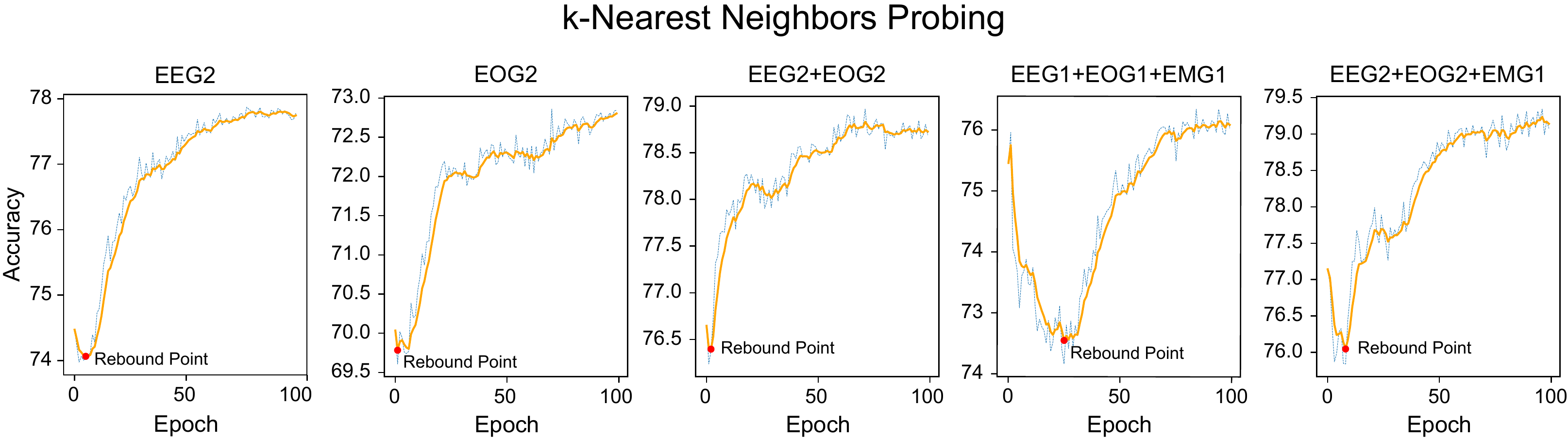}
\caption{Performance of k-nearest neighbors probing across training epochs for SynthSleepNet with various modality combinations.}
\label{fig:figure4}
\vspace{-5mm}
\end{figure*}

Figure \ref{fig:figure4} depicts the k-nearest neighbor (k-NN) probing performance during SynthSleepNet training. Like linear probing, k-NN probing was employed to evaluate the representational capacity of SSL. The output vectors were dimensionally reduced using principal component analysis and subsequently used as inputs for the k-NN algorithm.

The results revealed a distinct pattern, with performance declining in the initial stages of training but improving after a specific threshold, referred to as the “rebound point.” Further analysis revealed that the rebound point occurred earlier when similar modalities were employed as inputs and were delayed for combinations of dissimilar modalities. For instance, training on similar modalities, such as EEG2, resulted in an earlier rebound point, whereas combinations like “EEG1+EOG1+EMG1” and “EEG2+EOG2+EMG1” showed later rebound points. Notably, “EEG1+EOG1+EMG1” showed the latest rebound point at 25 epochs.

\subsection{Hypnograms} 
Figure \ref{fig:figure5} presents the prediction results for subject \#202054. The top row displays the labels assigned by sleep experts. The left column illustrates the predictions of SynthSleepNet with linear probing, while the right column shows the predictions of SynthSleepNet+TCM with fine-tuning. Detailed analysis revealed that applying TCM and fine-tuning produced predictions closely aligned with expert-labeled data. SynthSleepNet+TCM using “EEG2+EOG2+EMG1” achieved predictions nearly identical to those of the sleep expert, differing by only three labels.

\begin{table}[!htbp]
\caption{Performance of linear probing across different loss balance scales ($\alpha$). }
\label{table:table6}
\centering
\resizebox{0.35\textwidth}{!}{
\renewcommand{\arraystretch}{1.3}
{\scriptsize 
\begin{tabular}{llcc|cccclll} 
\cline{4-7}
 &  &  & \multirow{2}{*}{$\alpha$} & \multicolumn{3}{c}{\textit{Performance}}                 &  &  &  &   \\ 
\cline{5-7}
 &  &  &                   & \textit{ACC}   & \textit{MF1}   & \textit{K}    &  &  &  &   \\ 
\cline{4-7}
 &  &  & 0.1               & 80.61          & 72.57          & 0.74          &  &  &  &   \\
 &  &  & 0.5               & 81.69          & 73.54          & 0.75          &  &  &  &   \\
 &  &  & \textbf{1.0}      & \textbf{82.71} & \textbf{73.91} & \textbf{0.76} &  &  &  &   \\
 &  &  & 1.5               & 81.81          & 73.45          & 0.75          &  &  &  &   \\
 &  &  & 2.0               & 81.69          & 73.47          & 0.75          &  &  &  &   \\
\cline{4-7}
\multicolumn{9}{r}{\footnotesize *\textbf{best results} in each row are shown in bold.} \\
\multicolumn{9}{r}{\footnotesize *$K$ = \textit{Kappa}} 
\end{tabular}}
}
\vspace{-5mm}
\end{table}
\begin{table}[!htbp]
\centering
\caption{Performance of fine-tuning with a temporal context module.}
\label{table:table7}
\resizebox{0.4\textwidth}{!}{
\renewcommand{\arraystretch}{1.3}
{\scriptsize
\begin{tabular}{c|c|ccc} 
\hline
\multirow{2}{*}{\textit{Model}} & \multirow{2}{*}{\begin{tabular}[c]{@{}c@{}}\textit{Context}\\\textit{Length}\end{tabular}} & \multicolumn{3}{c}{\textit{Performance}}          \\ 
\cline{3-5}
                                &                                                                                             & \textit{ACC}   & \textit{MF1}   & \textit{K}  \\ 
\hline
LSTM                            & 20                                                                                          & 85.17          & 77.98          & 0.80           \\
Multi-Head
  Attention          & 20                                                                                          & 85.36          & 77.98          & 0.80           \\
LSTM
  + Multi-Head Attention   & 20                                                                                          & 85.82          & 78.96          & 0.81           \\
Ours
  (= Mamba)                & 10                                                                                          & 87.15          & 81.70          & 0.82           \\
\textbf{Ours (= Mamba)}         & 20                                                                                          & \textbf{89.07} & 82.98          & \textbf{0.85}  \\
Ours
  (= Mamba)                & 30                                                                                          & 88.79          & \textbf{83.02} & 0.85           \\
\hline
\multicolumn{5}{r}{\footnotesize *\textbf{best results} in each row are shown in bold.} \\
\multicolumn{5}{r}{\footnotesize *$K$ = \textit{Kappa}} 
\end{tabular}}
}
\vspace{-5mm}
\end{table}

\begin{table}[!htbp]
\centering
\caption{Efficiency of different temporal context modules.}
\label{table:table8}
\resizebox{0.48\textwidth}{!}{
\renewcommand{\arraystretch}{1.3}
{\scriptsize
\begin{tabular}{c|c|ccc} 
\hline
\multirow{2}{*}[-1.5ex]{\centering\arraybackslash \textit{Model}} &
\multirow{2}{*}[-1.5ex]{\centering\arraybackslash \begin{tabular}[c]{@{}c@{}}\textit{Context}\\\textit{Length}\end{tabular}} &
\multicolumn{3}{c}{\textit{Efficiency}} \\
\cline{3-5} 
 & & 
\begin{tabular}[c]{@{}c@{}}\textit{Parameters}\\(MB)\end{tabular} & 
\begin{tabular}[c]{@{}c@{}}\textit{Throughput}\\(samples/sec)\end{tabular} & 
\begin{tabular}[c]{@{}c@{}}\textit{Latency}\\(ms)\end{tabular} \\ 
\hline
LSTM                        & 20 & 148.51  & 2.01  & 497.51  \\
Multi-Head Attention        & 20 & 144.37  & 2.44  & 409.84  \\
LSTM + Multi-Head Attention & 20 & 152.55  & 1.61  & 621.12  \\
Ours (= Mamba)              & 20 & \textbf{140.08}  & \textbf{2.82}  & \textbf{354.61}  \\ 
\hline
\multicolumn{5}{r}{\footnotesize *\textbf{best results} in each row are shown in bold.} \\
\end{tabular}}
}
\vspace{-7mm}
\end{table}

\section{Discussion}

This study proposes SynthSleepNet and Mamba-based TCM to address the limitations of existing sleep analysis methodologies. The proposed approach demonstrated superior performance across three tasks: sleep stage classification, apnea detection, and hypopnea detection. The results underscore the importance of overcoming the single-task focus prevalent in several deep-learning-based sleep analyses, enabling a more comprehensive evaluation of sleep states. Furthermore, SynthSleepNet outperformed methodologies designed for single tasks (Table \ref{table:table2}). The ability of SynthSleepNet to analyze unlabeled PSG data presents an opportunity to accelerate sleep research and support the development of related healthcare solutions.

SynthSleepNet represents a novel multimodal hybrid SSL framework that integrates masked prediction and contrastive learning, effectively leveraging EEG, EOG, EMG, and ECG data to achieve high representation learning performance. The combined application of masked prediction and contrastive learning operates complementarily, enhancing stability and facilitating the learning of robust, high-level representations (Figure \ref{fig:figure3}). Consequently, SynthSleepNet outperformed state-of-the-art SSL methodologies (Table \ref{table:table2}) and maintained strong performance in semi-supervised learning scenarios, even with only 1\% or 5\% of the labeled dataset (Table \ref{table:table4}).

Incorporating Mamba-based TCM during the fine-tuning of the pretrained SynthSleepNet significantly improved performance (Table \ref{table:table3}). The design of the Mamba-based TCM \cite{ref32}—unlike commonly used RNNs \cite{ref13} or multihead attention mechanisms \cite{ref4, ref5}—was critical to achieving these improvements (Table \ref{table:table7}). In conclusion, SynthSleepNet combined with the Mamba-based TCM outperformed state-of-the-art supervised learning methods, requiring extensive labeled datasets (Table \ref{table:table3}). Moreover, the performance gap was pronounced under semi-supervised learning conditions (Table \ref{table:table4}).

Despite its overall robustness, SynthSleepNet exhibited relatively degraded performance in detecting hypopnea events under the semi-supervised condition (Table \ref{table:table4}). This trend was also observed in prior models such as SalientSleepNet \cite{ref9} and SleepFM \cite{ref23}, suggesting that this is a general limitation in current approaches rather than an issue specific to SynthSleepNet. One core reason lies in the physiological characteristics of hypopnea: unlike apnea, which involves a complete cessation of airflow and triggers strong EEG or autonomic responses, hypopnea typically involves only a partial reduction in airflow \cite{ref3}. As such, its physiological signature in EEG and EOG is more subtle and ambiguous, making it inherently harder to detect—especially in the absence of sufficient supervision or modality diversity \cite{malhotra2002obstructive}.

\begin{figure*}[!t]
\centering
\includegraphics[width=0.74\textwidth]{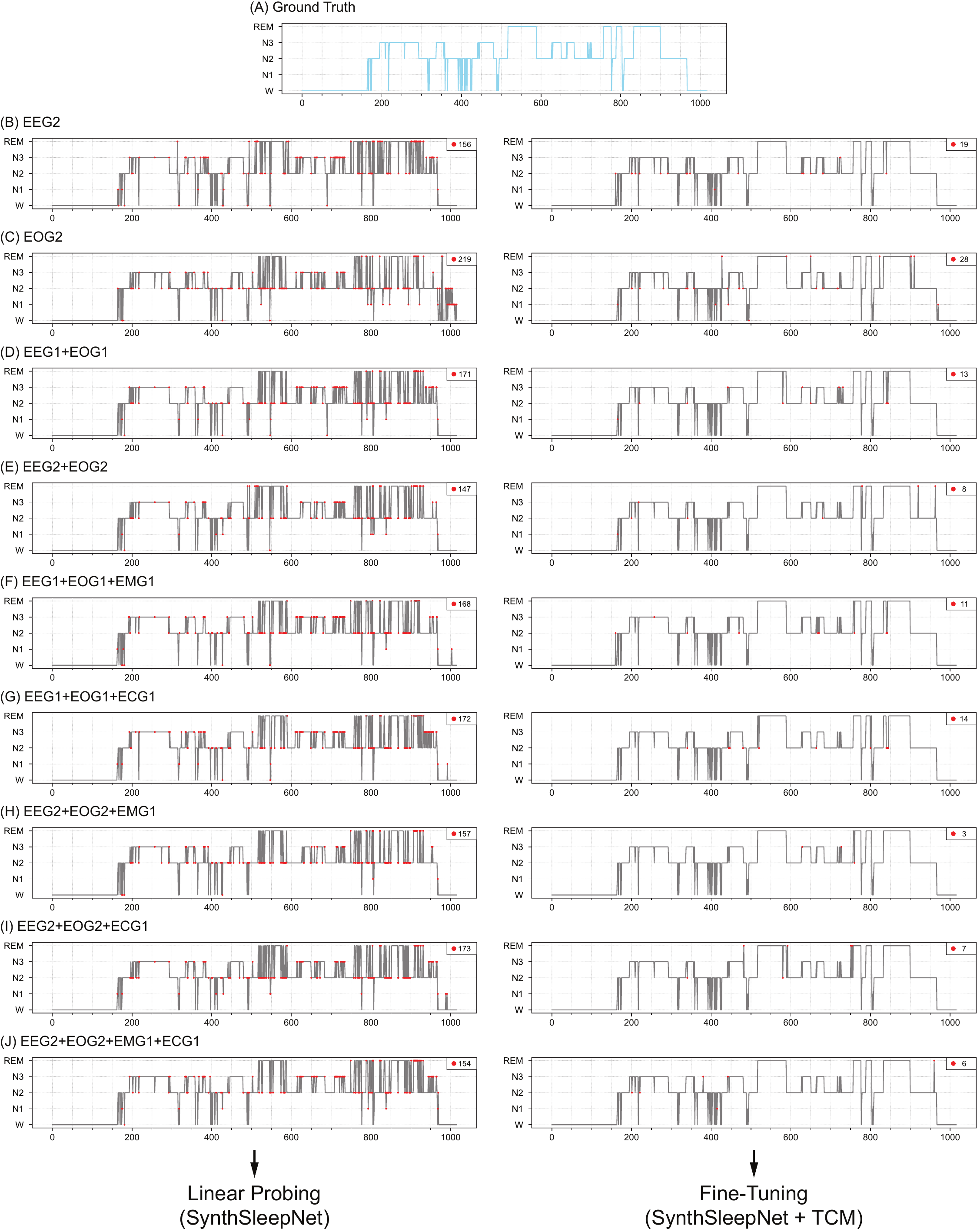}
\caption{Hypnogram for subject \#202054. (A) Expert-labeled sleep stage scoring. (B)–(J) The left column presents results from SynthSleepNet with linear probing, while the right column displays results from SynthSleepNet+TCM after fine-tuning. Errors are marked by red dots.}
\label{fig:figure5}
\vspace{-5mm}
\end{figure*}

Furthermore, the semi-supervised setting limits access to task-specific labels, reducing the model's ability to form explicit associations between weak signals and hypopnea events. This is particularly problematic for hypopnea, where physiological boundaries are vague and signal deviations are minimal. The absence of key respiratory-related modalities, such as airflow or thoracic/abdominal effort, further exacerbates the challenge \cite{redline2004effects}. Notably, when ECG—partially reflective of respiratory effort—was included, hypopnea detection performance improved markedly across all models. This reinforces the importance of integrating respiration-linked signals to improve model accuracy for subtle events. Future work may benefit from incorporating airflow or effort belts to further enhance hypopnea detection under weak supervision.

Recent mask-based SSL methodologies have demonstrated strong performance with mask ratios \cite{ref19, ref25, ref27} because higher mask ratios compel models to predict masked segments effectively, facilitating the learning of richer patterns and structures. However, SynthSleepNet operates with a relatively low mask ratio of 40\% despite employing a masked prediction task, which can be attributed to the following: First, excessively high mask ratios dilute the semantic information within the signal and fusion tokens, potentially impairing the contrastive learning capability of SynthSleepNet. Second, integrating multiple modalities increases the complexity of tasks of SynthSleepNet compared to single-modality methodologies. Consequently, a high mask ratio may overwhelm the model, leading to confusion.

Examining the performance of SynthSleepNet across various modality combinations revealed that EEG, EOG, and EMG produced the best results for sleep-stage classification (Tables \ref{table:table2} and \ref{table:table3}, Figure \ref{fig:figure5}), while ECG and EMG performed optimally for apnea and hypopnea detection (Tables \ref{table:table2} and \ref{table:table3}). This aligns with the guidelines outlined in the AASM manual \cite{ref24}, which is used by clinicians for sleep assessment. According to the AASM \cite{ref24}, EEG, EOG, and EMG are the key signals used for sleep-stage classification, while ECG, EMG, and airflow are essential for apnea and hypopnea detection. Combining all modalities (“EEG2+EOG2+EMG1+ECG1”) achieved high overall performance. However, it did not deliver the best outcomes for every task, likely due to the noisy and relatively limited nature of ECG data, which may cause distortion when integrated with other modalities.

The training process of SynthSleepNet exhibits a distinct “rebound point” phenomenon (Figure \ref{fig:figure4}), which deviates from typical patterns observed in deep learning models. This phenomenon reflects the time required for SynthSleepNet to effectively integrate information from different modalities. The “rebound point” occurred later when training on modalities with highly dissimilar features, indicating initial difficulty in reconciling modality discrepancies. Monitoring the “rebound point” using k-NN probing during training and adjusting training epochs accordingly is critical for optimizing performance.

SynthSleepNet has certain limitations despite several advantages. The relatively low \textit{Kappa} score for hypopnea detection highlights the need for improved data and labeling quality. Expanding dataset diversity is essential to enhance the generalizability of the model. Additionally, incorporating modalities such as photoplethysmogram signals, respiration signals, and sleep sounds may further improve performance. Third, additional sleep-related downstream tasks (e.g., arousal detection, SpO\textsubscript{2} desaturation detection, bruxism detection) should be explored for more comprehensive sleep analysis. Finally, lightweight optimization is essential to enable real-time processing in clinical applications.

SynthSleepNet has certain limitations despite its advantages. First, the performance of hypopnea detection was relatively lower compared to other downstream tasks, particularly when using EEG or EOG signals. This is because of the requirement of signals related to the heart and respiration, such as ECG and respiratory effort. This is evident in improved performance when using ECG, as observed from Table IV. Second, integrating additional physiological signals, such as photoplethysmogram, respiration signals, and sleep sounds, could further enhance model performance. Specifically, respiratory signals (airflow, respiration belt signal) are crucial for improving the accuracy of hypopnea and apnea detection. By incorporating multiple modalities, we anticipate the improvement of accuracy across various downstream tasks and hope to overcome the limitations of single-modality approaches. Third, in addition to the three downstream tasks performed in this study, we plan to expand by including additional sleep-related tasks, such as arousal detection, SpO2 saturation detection, and bruxism detection. This multi-task approach is expected to enhance the accuracy of sleep quality assessment and contribute to a more comprehensive monitoring of various sleep states.

\section{Conclusion}
This study introduces SynthSleepNet, a multimodal hybrid SSL framework, and Mamba-based TCM to overcome the limitations of existing deep learning methodologies for sleep assessment. SynthSleepNet integrates masked prediction and contrastive learning to effectively extract and fuse features from multimodal physiological signals (e.g., EEG, EOG, EMG, and ECG), facilitating the learning of high-level representations of PSG data. The Mamba-based TCM further improves model performance by capturing temporal dependencies within the PSG data. SynthSleepNet and Mamba-based TCM achieved superior performance in sleep-stage classification, apnea detection, and hypopnea detection while significantly reducing dependence on large-scale labeled datasets, as validated by experimental results. The proposed methodologies establish a robust foundation for advancing sleep research and broader applications in physiological signal analysis.

\appendices
\section{Training Setting and Hyperparameters}
The model training and evaluation were conducted on a computer equipped with an Intel I9-9980XE CPU (3.00GHz), 128GB RAM, and an NVIDIA 3090 GPU. All data processing and algorithm development were implemented in Python 3.10 using the PyTorch 2.0.1 library. Additionally, the hyperparameters used for NeuroNet, SynthSleepNet, and the downstream tasks are presented in Appendix Tables \ref{table:appendix_table1}, \ref{table:appendix_table2}, and \ref{table:appendix_table3}.

\setcounter{table}{0}
\begin{table}[!htbp]
\centering
\caption{Hyperparameters for NeuroNet.}
\label{table:appendix_table1}
\resizebox{0.25\textwidth}{!}{
{\scriptsize
\begin{tabular}{c|c} 
\hline
\textit{Hyperparameter} & \textit{Values}                  \\ 
\hline
epoch                   & 50\textbf{\textit{}}             \\
batch size\textbf{}     & 1024                             \\
frame size\textbf{}     & 3\textbf{\textit{}}              \\
overlap step\textbf{}   & 0.75\textbf{\textit{}}           \\
encoder dim\textbf{}    & 768\textbf{\textit{}}            \\
encoder depth\textbf{}  & 4\textbf{\textit{}}              \\
encoder
  head          & 8                                \\
decoder dim\textbf{}    & 256\textbf{\textit{}}            \\
decoder depth\textbf{}  & 3\textbf{\textit{}}              \\
decoder
  head          & 8                                \\
projection
  hidden     & {[}1024, 512]\textbf{\textit{}}  \\
temperature
  scale     & 0.05\textbf{\textit{}}           \\
mask
  ratio            & 0.75                             \\
optimizer               & AdamW                            \\
optimizer
  momentum    & (0.9,
  0.999)                   \\
learning
  rate         & 2e-05\textit{}                   \\
\hline
\end{tabular}}
}
\end{table}
\begin{table}[!htbp]
\centering
\caption{Hyperparameters for SynthSleepNet.}
\label{table:appendix_table2}
\resizebox{0.25\textwidth}{!}{
{\scriptsize
\begin{tabular}{c|c} 
\hline
\textit{Hyperparameter}         & \textit{Values}                 \\ 
\hline
epoch                           & 100                             \\
batch size\textbf{}             & 512                             \\
multimodal encoder dim\textbf{} & 512                             \\
multimodal
  encoder depth      & 4                               \\
multimodal
  encoder head       & 8                               \\
decoder dim\textbf{}            & 256\textbf{\textit{}}           \\
decoder depth\textbf{}          & 3\textbf{\textit{}}             \\
decoder
  head                  & 8                               \\
projection
  hidden             & {[}512, 256]\textbf{\textit{}}  \\
temperature
  scale             & 0.1\textbf{\textit{}}           \\
mask
  ratio                    & 0.4                             \\
lora\_r                         & 4                               \\
lora\_alpha                     & 16                              \\
lora\_dropout                   & 0.05                            \\
optimizer                       & AdamW                           \\
optimizer
  momentum            & (0.9,
  0.999)                  \\
learning
  rate                 & 0.0001\textit{}                 \\
\hline
\end{tabular}}
}
\end{table}
\begin{table}[!t]
\centering
\caption{Hyperparameters for Downstream Task.}
\label{table:appendix_table3}
\resizebox{0.5\textwidth}{!}{
{\scriptsize
\begin{tabular}{c|ccc} 
\hline
\textit{Hyperparameter}   & \textit{Evaluation Senario1} & \textit{Evaluation Senario2} & \textit{Evaluation Senario3}  \\ 
\hline
epoch                     & 150                          & 50                           & 50                            \\
batch
  size              & 512                          & 128                          & 128                           \\
optimizer                 & AdamW                        & AdamW                        & AdamW                         \\
optimizer
  momentum      & (0.9,
  0.999)               & (0.9,
  0.999)               & (0.9,
  0.999)                \\
learning
  rate           & 0.0002                       & 0.00025                      & 0.00025                       \\
temporal
  context length & -                            & 20                           & 20                            \\
mamba
  layers            &                              & 2                            & 2                             \\
mamba
  d\_state          &                              & 16                           & 16                            \\
mamba
  d\_conv           &                              & 4                            & 4                             \\
mamba
  d\_expand         &                              & 2                            & 2                             \\
\hline
\end{tabular}}
}
\end{table}

\begin{figure*}[!t]
\centering
\includegraphics[width=1\textwidth]{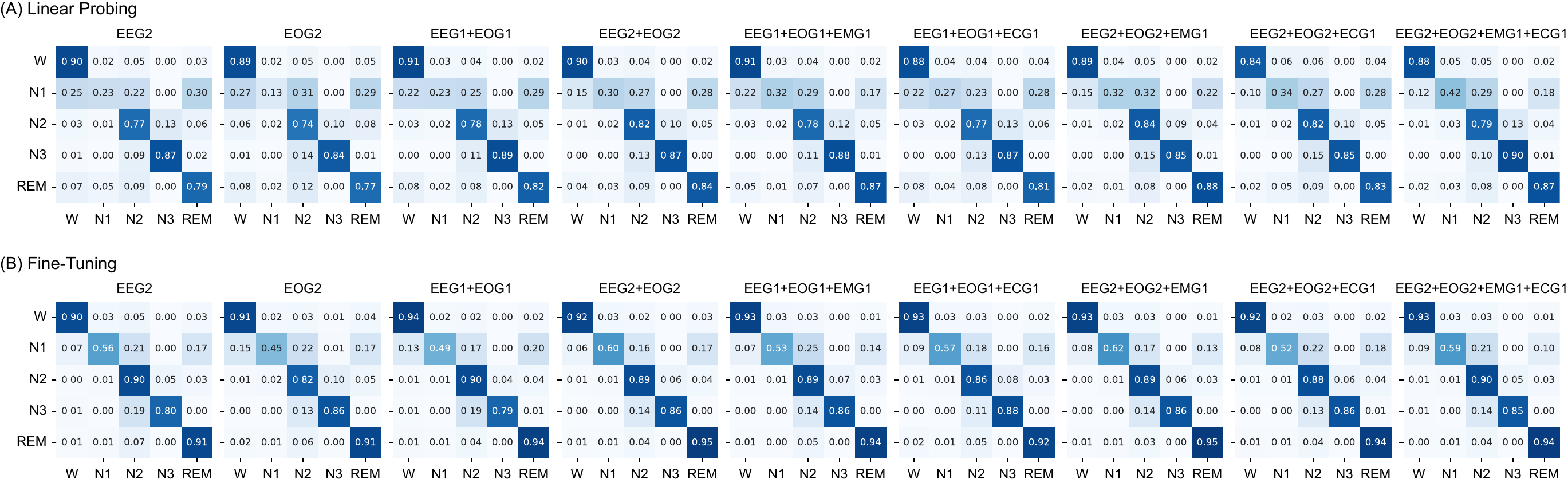}
\caption{Normalized confusion matrices for sleep stage classification using various combinations of EEG, EOG, EMG, and ECG signals under two evaluation settings: (A) linear probing and (B) fine-tuning.}
\label{fig:appendix_figure1}
\vspace{-5mm}
\end{figure*}

\section{Normalized Confusion Matrix for Sleep Stage Classification}
Appendix Figure \ref{fig:appendix_figure1}~(A) and (B) show the normalized confusion matrices for linear probing and fine-tuning, respectively. A comparison of the two indicates that fine-tuning results in an overall improvement in classification performance. Moreover, input configurations that incorporate multiple physiological signals consistently outperformed single-channel inputs, suggesting that integrating multimodal information facilitates more accurate sleep stage classification.

Among all stages, N1 exhibited the lowest classification performance, likely due to its ambiguous physiological characteristics and indistinct boundaries with adjacent stages such as Wake and N2. In the fine-tuning setting, the highest N1 classification accuracy (62\%) was achieved using the ``EEG2+EOG2+EMG1" combination, whereas the best result under linear probing (42\%) was observed with ``EEG2+EOG2+EMG1+ECG1". These findings suggest that EMG provides critical information for N1 classification, while ECG may, in some cases, introduce redundancy or noise.

\bibliographystyle{IEEEtran}
\bibliography{reference}

\begin{IEEEbiography}[{\includegraphics[width=1in,height=1.25in,clip,keepaspectratio]{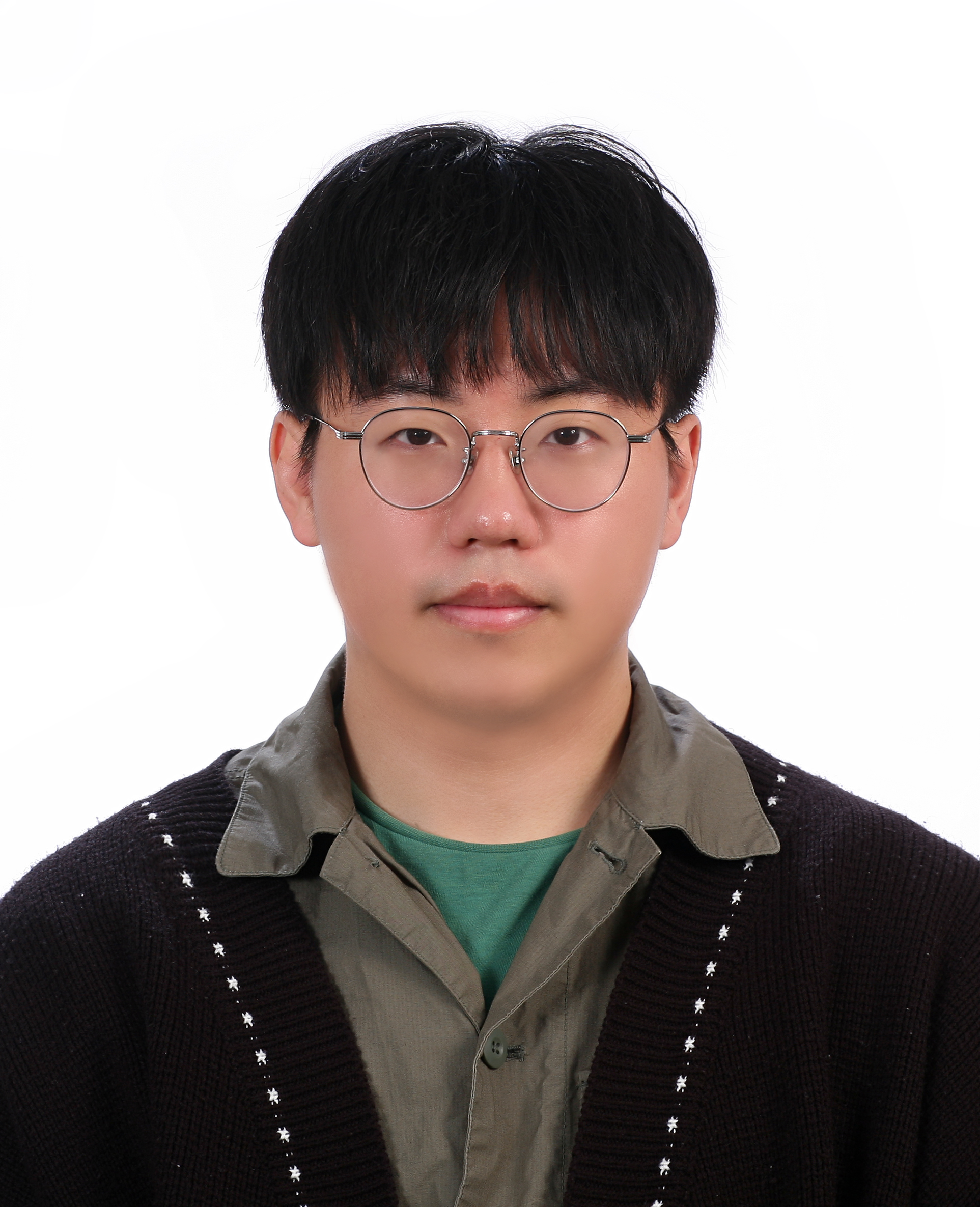}}]{Cheol-Hui Lee} is currently pursuing the Ph.D. degree with the Department of Brain and Cognitive Engineering, Korea University, Seoul, South Korea. His research interests include medical artificial intelligence and brain–computer interface.
\end{IEEEbiography}

\begin{IEEEbiography}[{\includegraphics[width=1in,height=1.25in,clip,keepaspectratio]{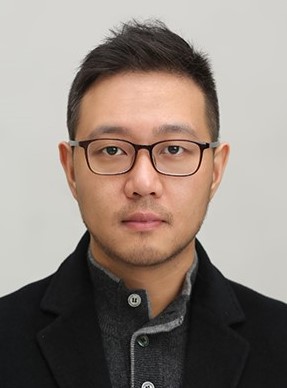}}]{Hakseung Kim} received the Ph.D. degree in brain and cognitive engineering from Korea University, Seoul, South Korea. He has been a Research Professor with the Department of Brain and Cognitive Engineering, Korea University, since 2018. His research interests include patient monitoring in the critical care environment and cerebrospinal fluid dynamics.
\end{IEEEbiography}

\begin{IEEEbiography}[{\includegraphics[width=1in,height=1.25in,clip,keepaspectratio]{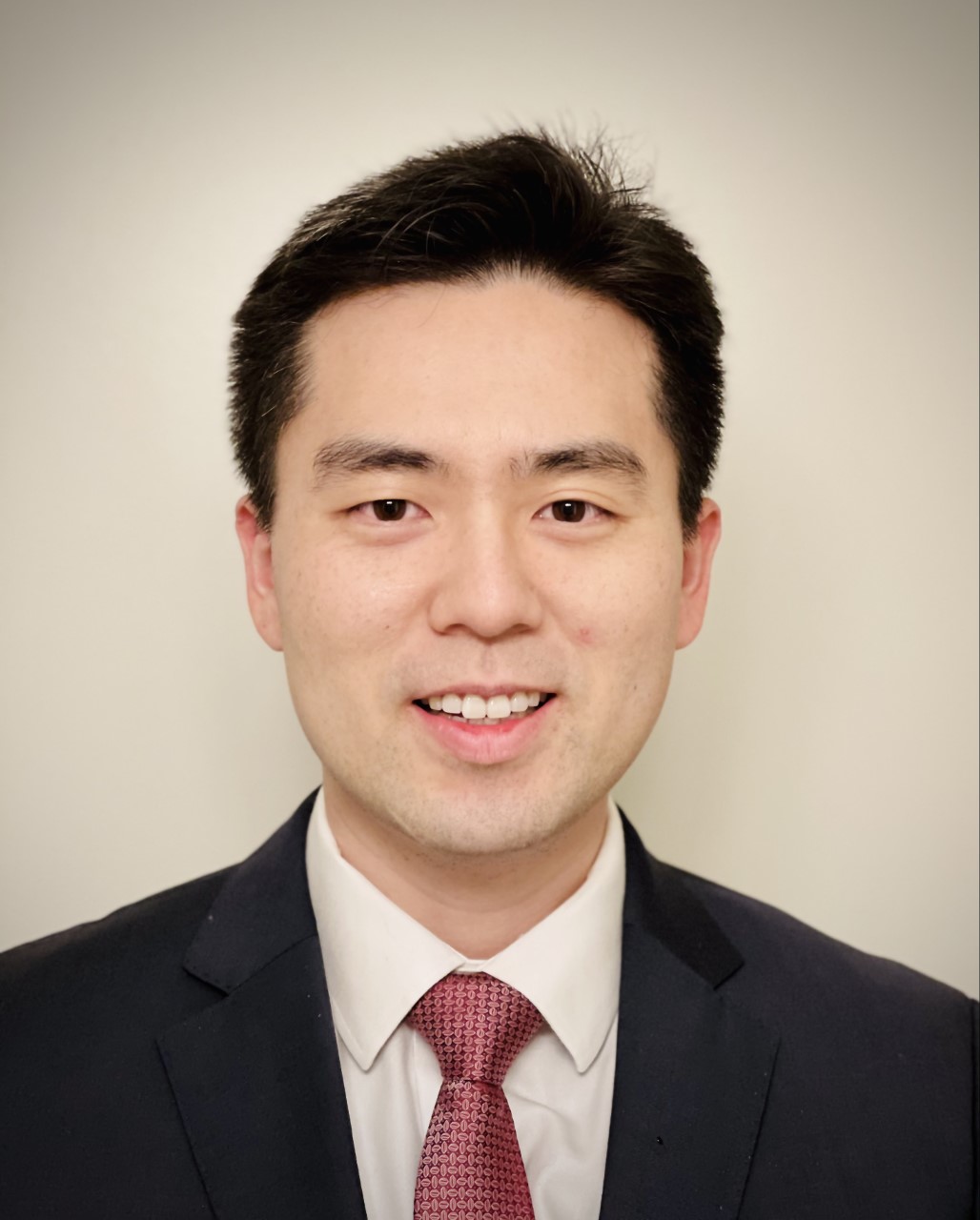}}]{Byung Chul Yoon} is an Assistant Professor of neuroradiology at the Stanford School of Medicine and an attending neuroradiologist at Veterans Affairs Palo Alto. He received an MD from the Johns Hopkins School of Medicine and a PhD in Physiology, Development, and Neuroscience at Cambridge University, UK. He completed a diagnostic radiology residency at Stanford Hospital and Clinics and a neuroradiology fellowship at Massachusetts General Hospital/Harvard Medical School. Following the fellowship, he stayed on at Massachusetts General Hospital/Harvard Medical School as a neuroradiology faculty, before returning to Stanford in 2022. Dr. Yoon’s research interests include neuroimaging, neuromodulation and translation of artificial intelligence into clinical practice. 
\end{IEEEbiography}

\begin{IEEEbiography}[{\includegraphics[width=1in,height=1.25in,clip,keepaspectratio]{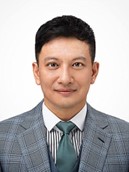}}]{Dong-Joo Kim} received the Ph.D. degree in engineering from the University of Cambridge, U.K. His time was split between clinical research (in the neurosurgery unit at the Addenbrooke’s Hospital) and brain modeling (the Engineering Department), University of Cambridge. Afterwards, he spent two years as a Research Fellow at the Neuroscience Mental Health Program, Department of Critical Care Medicine, Hospital for Sick Children, University of Toronto, Canada. Since 2017, he has been a Professor with the Department of Brain and Cognitive Engineering, Korea University, South Korea, and an Adjunct Professor with the Department of Neurology, Korea University College of Medicine, Seoul, South Korea. His research interests include brain engineering, artificial intelligence, and neuromodulation. 
\end{IEEEbiography}

\end{document}